\documentclass{elsarticle}
\usepackage[utf8]{inputenc}
\usepackage{graphicx}
\usepackage{amsfonts}
\usepackage{amssymb}
\usepackage{mathtools}
\usepackage{subcaption}
\usepackage{xcolor}
\usepackage{url}
\usepackage{soul}
\journal{ }
\newcommand{\bsym}[1]{\boldsymbol{#1}}
\usepackage{algorithm}
\usepackage{algpseudocode}
\title{Physics- and data-driven Active Learning of neural network representations for free energy functions of materials from statistical mechanics}
\author[umap]{J. Holber}
\author[usc]{K. Garikipati\corref{mycorrespondingauthor}}
\address[umap]{Applied Physics, University of Michigan}
\address[usc]{Department of Aerospace and Mechanical Engineering, University of Southern California}

\begin{document}

\begin{abstract}
Accurate free energy representations are crucial for understanding phase dynamics in materials. We employ a scale-bridging approach to incorporate atomistic information into our free energy model by training a neural network on DFT-informed Monte Carlo data. To optimize sampling in the high-dimensional Monte Carlo space, we present an Active Learning framework that integrates space-filling sampling, uncertainty-based sampling, and physics-informed sampling. Additionally, our approach includes methods such as hyperparameter tuning, dynamic sampling, and novelty enforcement. These strategies can be combined to reduce MSE—either globally or in targeted regions of interest—while minimizing the number of required data points. The framework introduced here is broadly applicable to Monte Carlo sampling of a range of materials systems.

\end{abstract}

\maketitle

\section{Introduction}

In recent years, machine learning (ML) models have proven to be increasingly valuable in disciplines such as image classification \cite{tharwat2023survey}, accelerated material discovery \cite{lookman2019active},  and training surrogate models \cite{tharwat2023survey,echard2011ak}.  As is now well understood, well trained ML model can provide insight to unlabeled or out of network data, without having to do time intensive experiments or use computationally expensive forward models.  

Our work is motivated by the continuum-scale modeling of materials physics, specifically using phase field methods to understand the phase dynamics and degradation effects of battery materials. Predictive phase field modeling of these processes relies on an accurate representation of the free energy density function. Historically, these free energies have been phenomenologically based which often can miss detailed physics such as high order interactions and  complex phase dynamics needed to match with and explain experiments.

With this background, we aimed to develop a framework enabling the creation of an atomistically informed free energy representation. The free energy densities can have compositions, order parameters, strains and temperature as arguments, attaining complex forms in the associated high-dimensional spaces. Therefore, we have employed  ML models--specifically neural networks--in workflows that bridge first principles statistical mechanics and continuum scale models to represent the free energy density \cite{shojaei2024bridging}. As an overview, our workflow uses density functional theory (DFT)-informed Monte Carlo (MC) to yield training data on generalized order parameter $\vec{\eta}$ and chemical potentials $\vec{\mu}$ pairs. As $\mu_i = \frac{\partial g}{\partial\eta_i}$  where $g$ is the free energy density, we can use an Integrable Deep Neural Network (IDNN) \cite{shojaei2024bridging,Teichert2019,Teichert2020} to recover the free energy density from the derivative data. A major source of error in this process is in choosing ideal sampling points for training our free energy surrogate model, due to the high-dimensionality of the system, and is the focus of this paper.

ML models, such as those of high-dimensional free energy densities, offer several advantages. Once trained, they are generally computationally inexpensive to evaluate, allowing for extensive design space exploration and analysis. They are nonlinear input/output maps, which importantly, can be integrable or differentiable-yielding important mathematical properties like the Hessian. This provides greater insight to the system, including optimization operations, and the identification of energy wells/regions of phase instability.

Two key, if obvious, aspects of ML include the type of model and the selection of data for training. When considering what data points to sample the ``one-shot'' method is a simple approach to training an ML model: A test matrix is constructed using a space-filling method (e.g., grid sampling or Latin Hypercube sampling), and the underlying model is evaluated at each point to generate input-output pairs. The ML model is then created and fitted to these pairs. If performance criteria are not met, the data and model are discarded, and the process repeats with a larger test matrix. This method, while straightforward, can lead to oversampling and poor representation in complex regions \cite{eason2014adaptive}.

Sequential sampling, a common alternative, begins with an estimated lower bound on the number of sample points and gradually adds new ones until a desired accuracy is achieved. The choice of these points is the core of an Active Learning algorithm. Generally, an acquisition function, or a set of acquisition functions, determines new points that are informative enough to create an accurate  model, especially in regions of interest, while minimizing calls to the fundamental model. Physical systems, including free energy density functions often have sub-domains that are more complex or physically important, requiring dense sampling. Oversampling these regions, however, can result in a model overtrained to these sub-domains and not generally applicable.

A combination of explorative (global/space-filling) sampling and exploitative (adaptive) sampling is often used. Explorative sampling provides a general overview of the parameter space, identifying regions of complexity, while exploitative sampling refines these regions. An important goal for Active Learning methods is not only having a good representation for the points that the model is trained on, but also being able to predict, with some accuracy, out-of-distribution points. While Active Learning focuses on iteratively improving the model by selecting the most informative samples, there remains a challenge in ensuring that the model can generalize well to unseen data. This involves maintaining a balance between exploiting known regions where the model performs well and exploring new, uncharted areas of the input space. The ultimate goal is to build a model that not only excels on the training data but also demonstrates robustness and reliability when confronted with novel inputs.

Various space-filling methods can be utilized to ensure a comprehensive exploration of the input space in Active Learning with neural networks. These methods include grid search (factorial design), Latin Hypercube Sampling (LHS), Delaunay triangulation/Voronoi tessellation, Sobol sequences, Halton sequences, Hammersley sequences, and Billiard walk sampling \cite{eason2014adaptive,wu2023comprehensive,metta2021novel,Polyak2014}. Grid search, or factorial design, is straightforward but can be computationally expensive, especially in high-dimensional spaces. LHS is more efficient, ensuring that sample points are spread evenly across the entire input space, and thus reducing the chances of clustering. Delaunay triangulation and Voronoi tessellation are geometric methods that provide structured techniques to fill the space by dividing it into regions based on proximity to sample points. Sobol, Halton, and Hammersley sequences are quasi-random methods that generate low-discrepancy sequences, leading to more uniform coverage of the input space compared to purely random sampling. These sequences are particularly useful in high-dimensional spaces where traditional grid search becomes impractical. Billiard walk sampling, a more recent technique, has specific advantages in maintaining uniformity and avoiding the clustering of points. It works by simulating the movement of a particle within any specified bounds, ensuring that the entire space is explored without bias towards any region. The choice of global sampling method depends on the specific requirements of the problem at hand, such as the dimensionality and geometry of the input space, and the computational resources available. 

Exploitative sampling methods are more varied and subject-dependent. Most use some error-based approach. For instance, in Physics-Informed Neural Networks (PINNs), additional points are sampled near previously trained points with high residuals \cite{wu2023comprehensive}. For other neural networks or Kriging models,  often the Mean Squared Error (MSE) between the model prediction and the training label is evaluated, and additional points are sampled nearby. High-variance regions, where the model differs significantly under different hyperparameters or data subsets, also warrant additional sampling. Methods like Query by Committee  (QBC), jackknifing, and prediction switches are used to explore these differences \cite{metta2021novel}. For problems with discrete outputs (e.g., positive/negative), sampling near decision boundaries is beneficial. For continuous outputs, sampling in highly sensitive regions, where small input changes lead to significant output changes, is advantageous. Some problems involve a finite set of available data points (e.g., in medical imaging, where the dataset is limited to collected scans), while others can involve a bounded space, with an infinite number of points within that space. The importance of different regions in the input space can also vary depending on the usage of an ML model, with some regions being more critical for the problem at hand. In this work, we implement our Active Learning approach, including a set of acquisition functions, for sampling using DFT-informed Monte Carlo simulations with the CASM (A Cluster's Approach to Statistical Mechanics) code \cite{casm}.

\subsection{Background/Motivation}

The free energy density function plays a fundamental role in materials physics, ranging over phenomena of chemistry, mass and heat transport, as well as mechanics \cite{callen1991thermodynamics,porter2021phase,garikipati2006continuum}. The free energy of materials systems that feature coupling between these phenomena  exists in a high dimensional space of composition, order parameters, temperature and strain. We have previously developed a scale bridging method to perform phase field simulations using a first principles statistical mechanics-informed free energy \cite{shojaei2024bridging}, as opposed to the conventional phenmonological free energy. We studied  the phase stability of Lithium Cobalt Oxide (LCO), a widely used cathode material in batteries, using phase filed simulations. In the context of LCO's phase stability, the conserved composition and six non-conserved order parameters derived from symmetry group considerations are sufficient to fully characterize its ordering, where the order parameters quantify the ordered arrangements of intercalating Li atoms on the corresponding sub-lattice  (Section \ref{sec:order params methods}). This ordering appears around compositions of Lithium of $\frac{1}{2}$ as shown in Figure \ref{fig:phasediagram}. 

An overview of the methodology follows in  Sections \ref{sec:order params methods}-\ref{sec:IDNN}, but is brief because it is not the central focus of this communication. For details, the interested reader could consult Faghih Shojaei et al. \cite{shojaei2024bridging}. We perform Density Functional Theory (DFT) calculations with an on-site Hubbard $U$ term (DFT$+U$) and van der Waals functionals for a set of different LCO supercell sizes and Li compositions. The results from these DFT computations are used to parameterize a cluster expansion Hamiltonian for the formation energy $E_f$. The cluster expansion is used in Semi-Grand Canonical Monte Carlo simulations. We also determine orderings and their corresponding order parameters $\vec{\eta}$  as described in \ref{sec:order params methods}. These Monte Carlo simulations, as described in \ref{sec:MC}, yield datasets with $\vec{\eta}$ and chemical potentials $\vec{\mu}$ where the component-wise relations are: 
\begin{equation}
    \mu_i = \frac{\partial g}{\partial \eta_i}
\end{equation}
 for free energy density $g$. Therefore, the Monte Carlo simulations yield free energy derivative data for a given ordering as described by $\vec{\eta}$. An Integrable Deep Neural Network (IDNN) can be trained to the derivative data and then analytically integrated to recover the free energy density as a function of $\vec{\eta}$ as described in \ref{sec:IDNN}. As discussed above, the Active Learning workflow is required to ensure well-distributed as well as targeted sampling of the composition-order parameter space, which is important for an accurate free energy model using the IDNN. The resulting phase field calculations allow us to model phase dynamics of the order-disorder transition and charge cycling in terms of temperature, particle size, and morphology (including different ordered variants) \cite{shojaei2024bridging, Teichert2019, Teichert2020}.

\subsubsection{Symmetrically Equivalent Order Parameters}
\label{sec:order params methods}

\begin{figure}
\centering
\begin{subfigure}{0.3\linewidth}
         \centering
         \includegraphics[width=1.0\linewidth]{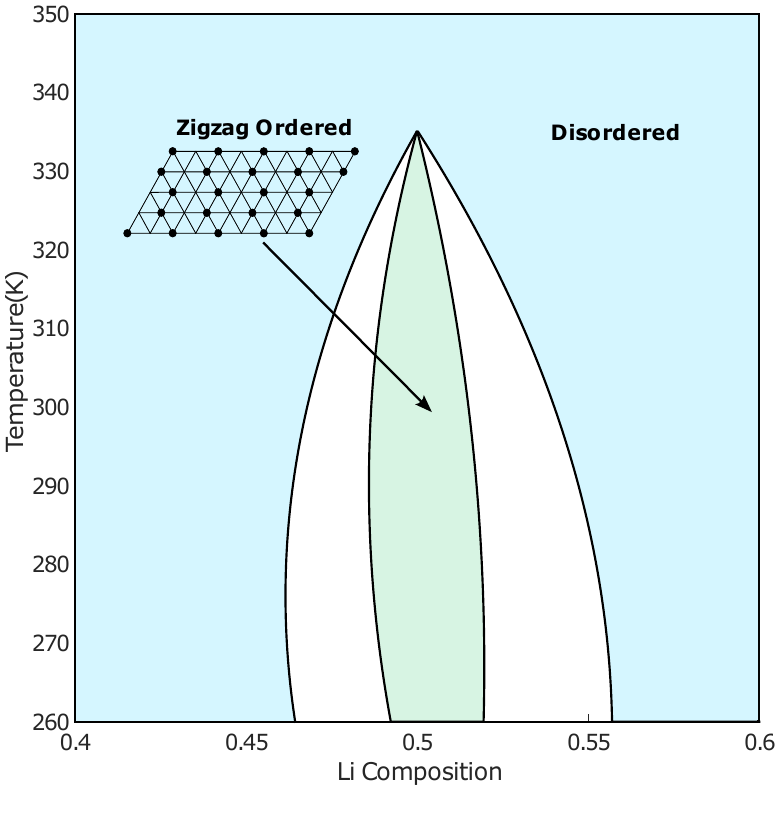}
         \caption{}
         \label{fig:phasediagram}
\end{subfigure}
\begin{subfigure}{0.65\linewidth}
         \centering
         \includegraphics[width=1.0\linewidth]{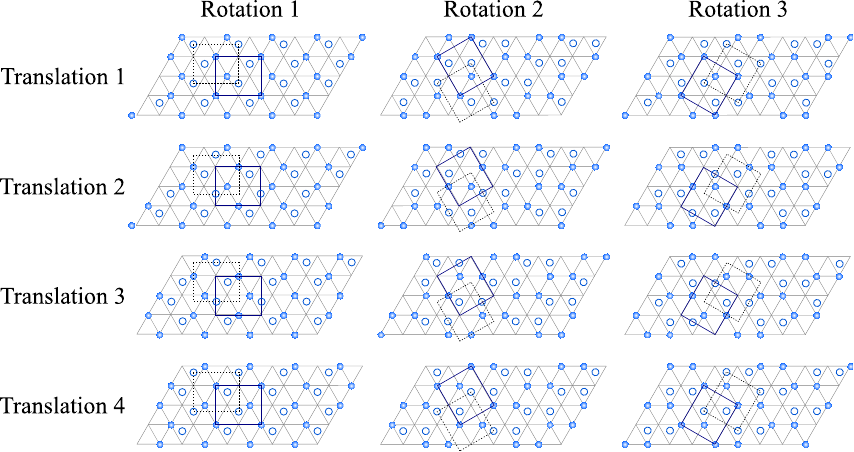}
         \caption{}
         \label{fig:variants}
\end{subfigure}
\caption{ (a) The phase diagram was constructed using a tangent line approach based on the 1D IDNN at various temperatures with zigzag ordering. (b)  The 12 variants corresponding to the zigzag ordering. The highlighted boxes on each variant show the smallest supercell which captures the details of the ordering. These three supercells were used to construct a 32-site mutually commensurate supercell \cite{teichert2023bridging}}
\end{figure}

The zig-zag ordering corresponding to the lowest free energy yields a set of 12 variants shown in Figure \ref{fig:variants}. The variants and their corresponding order parameters are determined following Natarajan et al. \cite{Natarajan2017}. The variants arise from three rotations and four translations belonging to the symmetry group on the triangular lattice. The mutually commensurate supercell which includes the supercells of all variants, is found to have 32 unique sublattice sites on two Li layers. The 32 sublattice sites are represented by a vector $\vec{x} = [x_1 x_2 .. x_{32}]$ where $x_i=1$ if the site is occupied and $x_i=0$ otherwise. 

The symmetry group of the zig-zag ordering can be represented by a P-invariant matrix following the algorithm in Ref. \cite{thomas2017}. The eigenvalue decomposition of P yields 32 distinct eigenvectors with degeneracies in eigenvalues corresponding to symmetrically equivalent variants. The eigenvectors form columns of a matrix $\boldsymbol{Q}$, such that $\vec{\eta} = \bsym{Q} \vec{x}$, where $\vec{\eta}$ represent the order parameters. The matrix $\bsym{Q} $ is of dimension $32\times 32$; however only the first 7 columns correspond to the zig-zag ordering so we are able to use a reduced matrix $\bsym{Q} \in \mathbb{R}^{7\times 32}$  in our calculations, where one row corresponds to the composition of Li, and the other 6 corresponding to orderings. Therefore, $\vec{\eta}$ is a vector of length seven, with $\eta_0$ corresponding to the composition and $\eta_1$ through $\eta_6$ representing different order parameters. Going forward, $\eta_i$ will denote one of these order parameters where $i$ can be between $1$ and $6$. Since the order parameters are invariant with respect to the free energy, the specific value of $i$ is inconsequential. An order parameter value of $\pm 0.5$ corresponds to a completely ordered variant. If all order parameters are 0, this represents a disordered system. This method allows us to define the order parameters $\vec{\eta}$ as ensemble averages over our Monte Carlo calculations, and to represent the 32-site mutually commensurate supercell with the seven order parameters.

\subsubsection{Monte Carlo}
\label{sec:MC}
\begin{figure}
    \centering
    \includegraphics[width=0.5\linewidth]{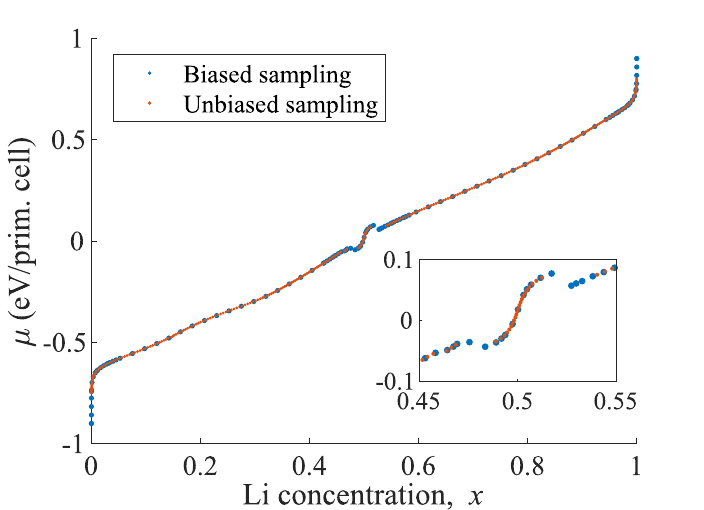}
    \caption{The results of Monte Carlo simulations varying Lithium composition at 300 K with and without bias sampling (umbrella sampling).}
    \label{fig:chem pot}
\end{figure}

We use the \texttt{CASM} code \cite{casm} to calculate the Li composition as a function of chemical potential using the Semi-Grand Canonical Ensemble Monte Carlo (initially ignoring orderings). Initial unbiased samplings show a gap in the composition values verifying the existence of an order-disorder transition region (a region of phase instability) as seen in Figure \ref{fig:chem pot}.  Phase field simulations require free energy information within those regions so we employ biased (umbrella) sampling using bias potentials  \cite{Torrie1977,Sadigh2012a,Sadigh2012b}, with the corresponding partition function: 
\begin{align}
    \Theta &= \sum_{\bsym{\sigma}} \exp{\left(-\frac{E(\bsym{\sigma})  + \sum_{j=0}^6\phi_j(\eta_j(\bsym{\sigma}) - \kappa_j)^2}{k_B T}\right)} \label{eqn:part3}
\end{align}
where $\phi_i$ and $\kappa_i$ determine the curvature and center of the bias potential, respectively, and the inner sum is over the composition and six order parameters (described in Section \ref{sec:order params methods}). 

The ensemble average of the composition $\langle \eta_0\rangle$ and each order parameter $\langle\eta_i\rangle$, $i = 1,\dots,6$  is related to its corresponding chemical potential through the bias parameters:
\begin{align}
    \frac{1}{M}\mu_i\Big|_{\langle\vec{\eta}\rangle} &= -2\phi_i(\langle\eta_i\rangle - \kappa_i), \qquad i=0,\ldots,6
    \label{eq:mu-eta-kappa}
\end{align}
where $M$ is the number of lattice sites. Using $\mu_i = \frac{\partial g}{\partial\eta_i}$
\begin{align}
    \frac{\partial g}{\partial \eta_i} \Big|_{\langle\vec{\eta}\rangle} = \mu_i \Big|_{\langle\vec{\eta}\rangle} = -2M\phi_i(\langle\eta_i\rangle - \kappa_i) \qquad i=0,\ldots,6
    \label{eq:g-mu-eta-kappa}
\end{align}

Given $E_\text{f}(\bsym{\sigma})$, we sample within the semi-grand canonical ensemble, in which the chemical potential is specified and the corresponding composition and/or order parameters are determined through ensemble averaging. 
The Monte Carlo calculations have a convergence criterion of $\langle\eta_i\rangle=3 \times 10^{-4}$ and a precision of Var$(\langle\eta_i\rangle - \kappa_i)$ and serve as input for training our free energy model.

\subsubsection{Symmetry-invariant linear combinations of order parameters}
The 12 variants are symmetrically equivalent in the sense that they have the same energy. This property needs to be represented in our free energy density model to ensure physically consistent predictions and to reduce the number of data points which need to be sampled. Therefore, instead of using the order parameters $\vec{\eta}$  directly as the input in the neural network, we construct polynomial functions, which enforce these symmetries of the order parameters in our model. These polynomial invariants serve as the input features, ensuring the learned free energy function remains invariant under symmetry transformations.

To achive this, we apply the Reynolds operator to monomials up ot the 6th order as described in Ref \cite{dresselhaus2007group}.The Reynolds operator is used to project functions onto the invariant subspace, enforcing symmetries of the order parameters and their products up to degree six.
\begin{equation}
 h(\vec{\eta}) = \sum_{\bsym{M}^{(\eta)}\in\mathcal{M}}f(\bsym{M}^{(\eta)}\vec{\eta})
\end{equation}
where for each $\bsym{M} \in P$ we have $\bsym{M}^{(\eta)} = \bsym{QM}$. 

\subsubsection{Integrable Deep Neural Network (IDNN)}
\label{sec:IDNN}

A standard Deep Neural Network (DNN) can be represented by a function $Y(\bsym{X},\bsym{W},\bsym{b})$, which is not always inherently integrable. Since we want to train the neural network to the derivative of free energy we use an IDNN, described in Ref \cite{Teichert2019}, and represented by:
\begin{align}
    \bsym{\widehat{W}},\bsym{\widehat{b}} = \underset{\bsym{W},\bsym{b}}{\mathrm{arg\,min}}\,\sum_{k=1}^n\mathrm{MSE}\left(\frac{\partial Y(\bsym{X},\bsym{W},\bsym{b})}{\partial X_k}\Big |_{\bsym{\widehat{X}}_\theta},\widehat{y}_{\theta_k}\right)
\end{align}
The IDNN representation for free energy is obtained by training on the symmetry functions defined in the previous sections as the input and the chemical potentials as the output. The optimized weights $\bsym{\widehat{W}}$ and biases $\bsym{\widehat{b}}$ are used with the IDNN function $\partial Y(\bsym{X},\bsym{\widehat{W}},\bsym{\widehat{b}})/\partial X_k$ to predict the chemical potential. For our problem, the feature (input) $\widehat{X}_\theta$ are the symmetry-invariant combination of order parameters $h(\vec{\eta})$ and the label (output) $\widehat{y}_\theta$ is the chemical potential from Monte Carlo $\vec{\mu}$. Therefore $\partial Y(\bsym{X},\bsym{\widehat{W}},\bsym{\widehat{b}})/\partial X_k$ is being trained to predict $\vec{\mu}$ where $\mu_i=\frac{\partial g}{\partial \eta_i}$. The same weights and biases are then used in its antiderivative DNN function $Y(\bsym{X},\bsym{\widehat{W}},\bsym{\widehat{b}})$ to predict the homogeneous free energy density ($g$).

\subsubsection{Active Learning}
This approach to developing a free energy model necessitates adaptive sampling of non-convex regions, extrema, boundaries of admissible regions, and high-error points within the seven-dimensional space of the free energy density. We previously used global sampling (Billiard walk), sampling near known extrema (wells, vertices, and endpoints) and high error sampling. However, using these methods required close to 100,000 data points and was not well sampled in some regions, including in regions of non-convexities which are of particular importance for phase field simulations. This motivated the current work on additional criteria for data sampling to yield better representation of the free energy density. In Section 2 we discuss the criteria for sampling and other aspects of our active learning workflow including hyperparameter search, dynamic reweighting of criteria, and novelty. 

We divide the active learning workflow into rounds. In each round, our algorithm recommends data points in the $\vec{\eta}$ space to sample. In round 0, only global sampling and known areas of interest criteria are used to recommend points. This is because all other methods are exploitative and require an already trained neural network. In the following rounds, all user-specified criteria are applied, except for Query by Committee, which is only applied on rounds following hyperparamter searches. Next the algorithm uses Monte Carlo to find the $(\vec{\eta},\vec{\mu})$ pairs. Then a neural network is trained on the $(\vec{\eta},\vec{\mu})$ pairs from the current and previous, with an additional hyperparameter search for the neural network architecture under defined conditions. In Section 3 we show the results of Active Learning runs for LCO. Each run varies based on the set of data sampling criteria we use and other tecniques to improve the process including dynamic reweighting and novelty.

\section{Overview of Active Learning Workflow} \label{sec: overview}

\begin{figure}
    \centering
    \begin{subfigure}{0.32\linewidth}
        \centering
        \includegraphics[width=\linewidth]{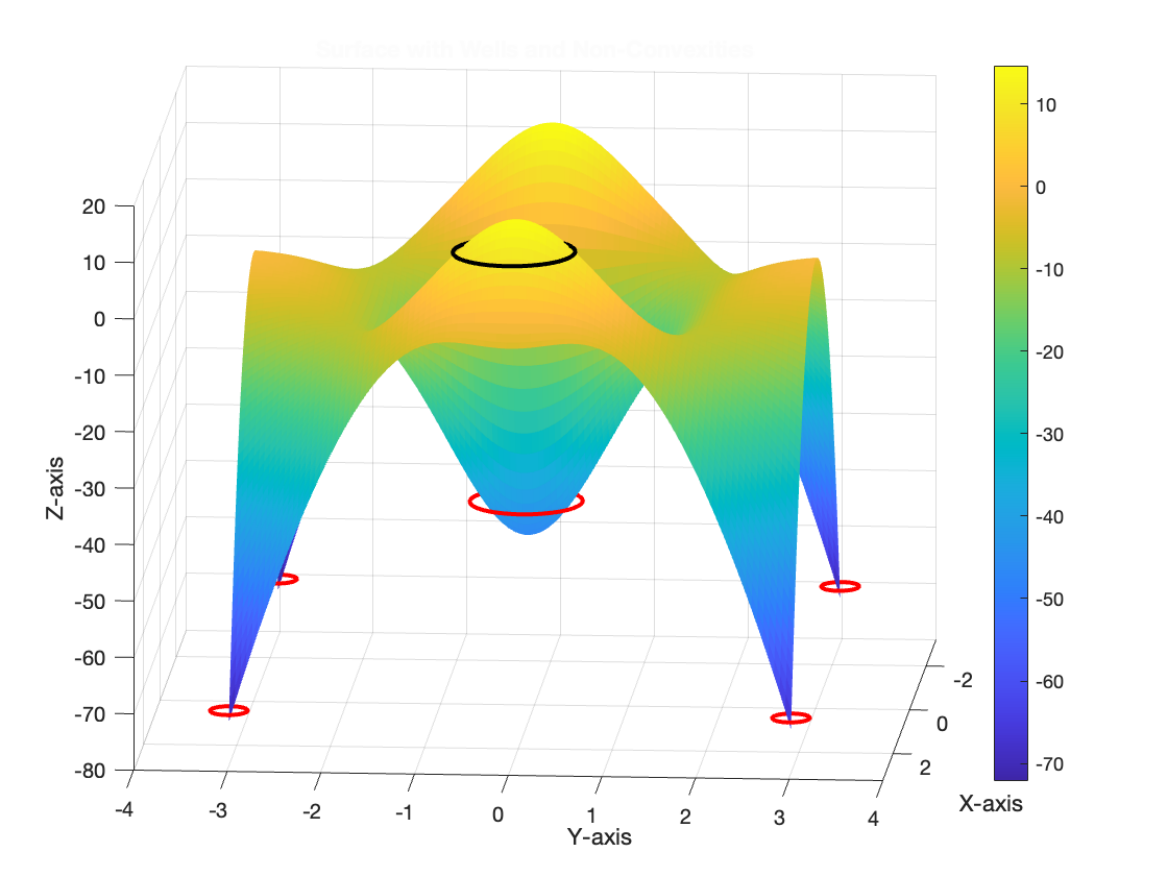}
        \caption{}
        \label{fig:sample surface}
    \end{subfigure}
    \hfill
    \begin{subfigure}{0.32\linewidth}
        \centering
        \includegraphics[width=\linewidth]{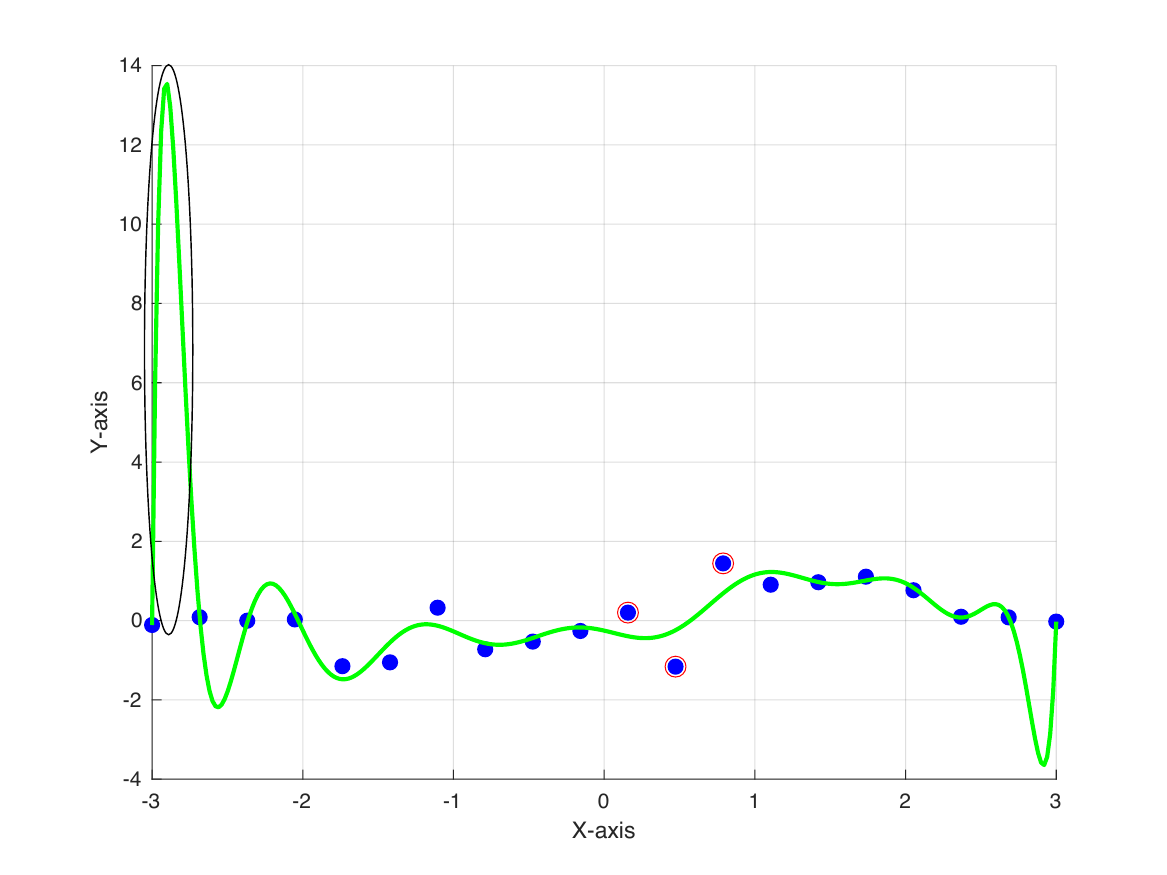}
        \caption{}
        \label{fig:error sensitivity}
    \end{subfigure}
    \hfill
    \begin{subfigure}{0.32\linewidth}
        \centering
        \includegraphics[width=\linewidth]{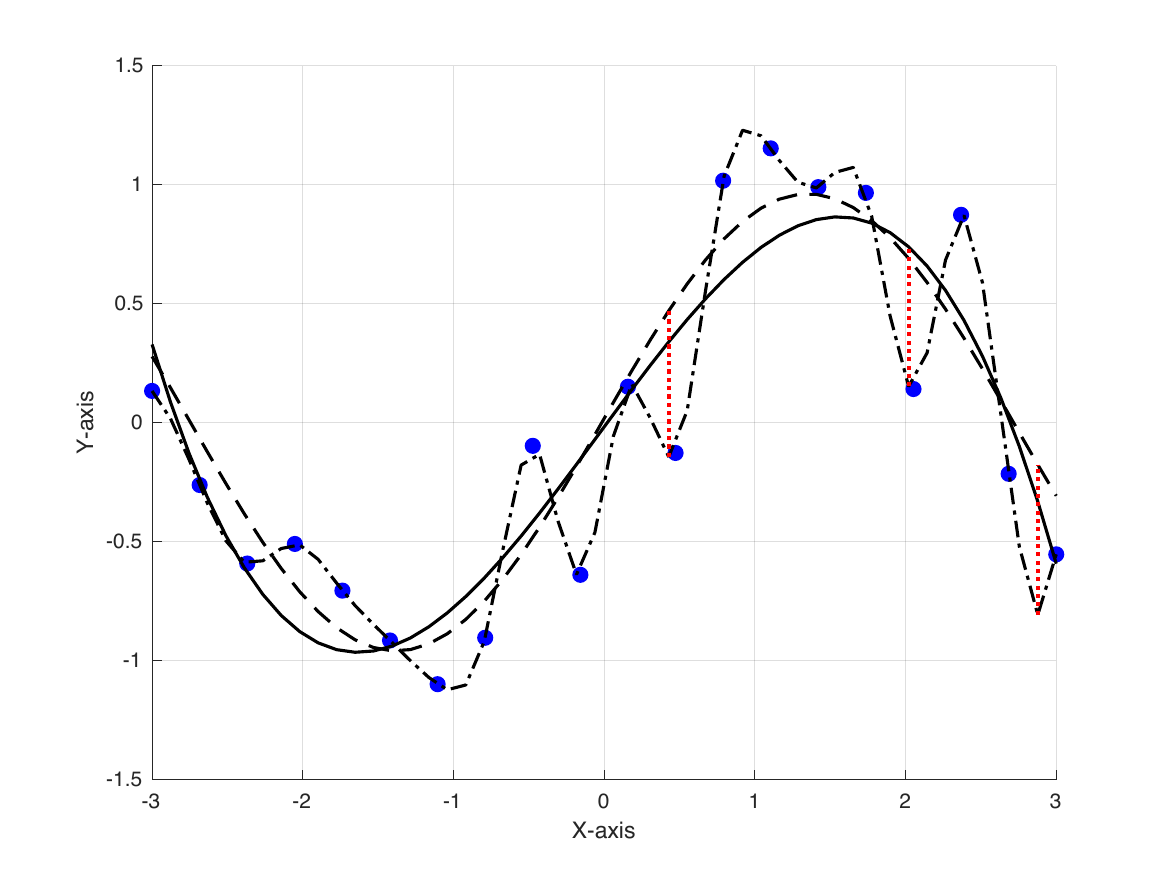}
        \caption{}
        \label{fig:qbc example}
    \end{subfigure}
    \caption{These figures show sample surfaces/data to illustrate regions of importance for our Active Learning workflow. (a) Illustration of a sample free energy surfaces. The energy wells are circled in red, and a region of non-convexity is circled in black. (b) A dataset (blue points) and its fit (green line). High-error points, where the fit is worst, are circled in red. A region with a large gradient (high sensitivity) is circled in black. (c) A dataset (blue points) with different fits (three curves). Query by Committee  samples additional points at the red dotted lines where the fits differ most. }
    \label{fig:combined_figure}
\end{figure}

\subsection{Criteria for Sampling}

Our active sampling method consists of a series of rounds, where for each round we sample data and train a neural network. We have implemented a variety of criteria for determining points to sample. The first two criteria discussed are both instances of Explorative sampling, ie not using the model results for guidance, consisting of global sampling with Billiard walk (\ref{global sampling Theory}) and known areas of interest sampling (\ref{KAE}). 

We also have implemented different Exploitative sampling methods, which use the results of the neural network trained during the previous round. We have implemented several general approaches to reducing errors in neural networks by sampling near points with High Error (Section \ref{High Error theory}), High Sensitivity (Section \ref{senstivity theory}), and large variations in model predictions (Section \ref{QBC theory}). We also have  implemented several methods which are important for modeling the physics of the problem including finding wells (Section \ref{find wells theory}), regions of phase instability (Section \ref{regions of instability}), and the path of lowest free energy near orderings (Section \ref{LFE theory}). 

The points considered for High Error sampling are based on the training data, and the path of lowest free energy is based on a previously defined set of points (detailed in Section \ref{LFE theory}). The rest of the Exploitative criteria use a testing set of data, generated using Billiard walk sampling. There are 1 million points generated by Billiard walk sampling and predicted using the neural network. Out of the million points, 95$\%$ are generated with Billiard walk sampling seven dimensions and 5$\%$ are generated using Billiard walk sampling with just the composition and one order parameter.

\subsubsection{Benchmark: global sampling} \label{global sampling Theory}

There are many methods for generating points uniformly in a region $Q \in \mathbb{R}^n$, including, but not limited to, such methods as grid sampling, uniform random sampling, Latin Hypercube, Sobol sequence, and Markov Chain Monte Carlo (MCMC) based sampling.  \cite{wu2023comprehensive}. One of the most well known MCMC based algorithms is Hit-and-Run (HR) \cite{Polyak2014}. While generally effective, hit-and-run algorithms often do not perform well in ``bad-shaped regions'' quoting Polyak et al\cite{Polyak2014}. In our work, which is high-dimensional bonded by a number of hyperplanes and and additionally it is unclear if there are non-convex regions, hit-and-run is not expected to perform well.

Billiard walk sampling is a type of random walk algorithm, combining Hit-and-Run with billiard trajectories. It is inspired by gas filling a vessel or the movement of billiard balls. This method ensures uniform coverage and minimizes clustering. It assumes that the particle moves with constant speed and is reflected at boundaries. The basic idea of the algorithm, described more in Algorithm \ref{alg:billiard}, is that given a position $x_i$, the next position $x_{i+1}$ is generated by choosing a random direction and length for the trajectory, and propagating the ``billiard ball'' in that direction, reflecting off boundaries when hitting them. The trajectory continues, changing direction with every reflection, until the specified length has been covered. 

We do Billiard walk sampling over the full seven dimensions of the composition and the six order parameters. This enables us to sample over the entirety of the input space, and due to the nature of Billiard walk sampling this tends to cluster away from the edges of the space. We also do Billiard walk sampling over two dimensions: the composition and one order parameter. This two-dimensional sampling is relevant because we expect the free energy at most compositions to be disordered, or close to ordered such that one order parameter dominates and the rest are close to 0. We pick $\eta_1$ for this sampling since the order parameters are symmetrically equivalent. Additionally, we perturb $\eta_2-\eta_6$ randomly by $\pm 0.05$ around 0. 
 
\begin{algorithm}
\caption{Billiard walk sampling}\label{alg:billiard}
\begin{algorithmic}[1]
    \Procedure{Billiard walk algorithm}{}
        \State Define a random starting position in the input space $x^0,i=0$
        \While{$i \leq N$}
            \State Define a trajectory direction ($\vec{d}$) and length ($l$)
            \State When the trajectory of the particle from $x_i$ ($\vec{d}$) meets a boundary with internal normal $\vec{s}$, $|s|=1$ set $\vec{d} \rightarrow \vec{d} - 2(\vec{d} ,\vec{s} )\vec{s}$, and record the location of a point near the boundary, but still within the input space
            \State The modified $\vec{d}$ is the new direction reflected off the boundary. It subtracts twice the component of d that aligns with s , effectively “flipping” the direction across the boundary normal
            \If{the particle does not meet  a nonsmooth boundary and the number of reflections is not greater than some value $R$} 
    \State Define the endpoint as $x_{i+1}$, and $i=i+1$
\Else
    \State Discard current trajectory 
    \State Note: We discard trajectories with too many reflections, as this likely means it is caught bouncing off the same boundaries and is not adequately sampling the reference space 
    \EndIf
        \EndWhile
        \State Return set of all points $x_0,x_1,...x_N$
    \EndProcedure
\end{algorithmic}
\end{algorithm}

For our work, the boundaries can be defined by constraining the composition of each sub-lattice site to satisfy $0 \le x_i \le 1$ , i.e., non-negative  with a maximum of $1$. Given the relationship  $\vec{\eta} = \bsym{Q}\vec{x}$  the boundaries are defined as:  

\begin{align}
    0 \leq \sum_{j=0}^6 Q^{-1}_{ij}\eta_j \leq 1, \qquad i = 0,\ldots,31
    \label{eq: constraints on boundary}
\end{align}

Following from  $\bsym{Q}$, in the case with only 1 nonzero order parameter, this reduces to: 

\begin{equation}
    |\eta_1| \leq 0.5-|(0.5-\eta_0)| 
\end{equation}

\subsubsection{Known Areas of Interest: Wells/End members/vertices} \label{KAE}

With knowledge about the properties of the crystal or free energy density surface, we can sample in ``Regions of Interest'' \cite{liu2018survey}. We want to use this information to guide our sampling. For LCO, from Monte Carlo simulations we know the energy wells occur at orderings ie $\eta_0 = 0.5$, $\eta_i = \pm 0.5$ for some $i$, and $\eta_j = 0$, for $j \neq i$ (Well 1).  There is also a local minimum at $\eta_0=0.5$ and $\eta_i=0,i=1,...,6$ (Well 2) and we sample additionally there. The algorithm is shown in more detail in Algorithm \ref{alg: sample wells}. These wells can be seen in Figure \ref{fig:eta1eta2}, where several figures, including the Active Learning runs corresponding to High Error, Query by Committee  (QBC), and Sensitivity, show both Well 1 and Well 2.

\begin{algorithm}
\caption{Sampling Wells Procedure} \label{alg: sample wells}
\begin{algorithmic}[1]
\Procedure{SampleWells}{rnd}
    \State Inputs: Perturbation magnitude($P$), Number of times to sample within each well ($N_w$), and the dimensions(dim). For LCO there are 7 dimensions, one corresponding to the composition of Lithium, and the other six correspond to order parameters each of with have two wells. 
    \State Initialize $\eta_w$ as a zero matrix of size $(2 \cdot$ \text{dim}-1, dim)
    \State Set $\eta_w[:, 0] \gets 0.5$ - This dimension corresponds to the composition of Lithium. Ordering occurs around this region, and we know wells occur when $\eta_0=0.5$ and $\eta_i=\pm0.5 (i=1,..\text{dim}-1)$ 
    \For{$i = 1$ to dim$-1$}
        \State $\eta_w[2i, i] \gets 0.5 - \frac{P}{2}$
        \State $\eta_w[2i+1, i] \gets - ( 0.5 - \frac{P}{2})$
        \State These dimensions correspond to order parameters, with $n_{w}[:,i] = \pm 0.5, i =1,..,\text{dim}-1$  being fully ordered
\State For the example of $P=0.05$, this creates a starting set of points for our wells exploration. For each order ordering (represented by an order parameter being $\pm 0.5$) we begin with the same initial location in the well. We begin at $0.5-P$ so that we can perturb throughout the well without going out of the boundary. 

\[
\begin{bmatrix}

0.5 & 0.45 & 0 &  \cdots  \\

0.5 & -0.45 & 0 & \cdots \\

0.5 & 0 & 0.45 & \cdots \\
\vdots & \vdots & \vdots & \cdots \\

\end{bmatrix}
\]

\EndFor

    \State Once we have the initialization points we replicate each row in $\eta_w$ $N_w$ times, ie if $N_w=3$ then the new matrix would look like the following. 
    \[
\begin{bmatrix}

0.5 & 0.45 & 0 &  \cdots  \\

0.5 & 0.45 & 0 &  \cdots  \\

0.5 & 0.45 & 0 &  \cdots  \\

0.5 & -0.45 & 0 &  \cdots  \\

\vdots & \vdots & \vdots & \cdots \\

\end{bmatrix}
\]
    \State  We then sample through the well region by perturbing each point in $\eta_W$ by $P$. Having the $\eta_i=0.5-P$ helps for the procedure to stay within the boundary plane defined be \ref{eq: constraints on boundary}. Therefore the final results would look something like the following (assuming $P=0.05$ and $N_w=3$, where each $|\delta| < P$.
        \[
\begin{bmatrix}

0.5+\delta_1 & 0.45+\delta_2 & 0+\delta_3 &  \cdots  \\

0.5+\delta_4 & 0.45+\delta_5 & 0+\delta_6 &  \cdots  \\

0.5+\delta_7 & 0.45+\delta_8 & 0+\delta_9 &  \cdots  \\

0.5+\delta_{10} & -0.45+\delta_{11} & 0+\delta_{12} &  \cdots  \\

\vdots & \vdots & \vdots & \cdots \\

\end{bmatrix}
\]
      
\EndProcedure
\end{algorithmic}
\end{algorithm}

For the case of LCO, we sample at a lithium composition of  $\eta_0 = 0.5$ while varying a single ordering parameter $\eta_i$ within the range  $\eta_i = -0.5$  to  $\eta_i = +0.5$, with all other parameters  $\eta_j = 0$  for  $j \neq i$. This approach is crucial because, in phase dynamics, the region of ordering is especially important. As defined, this corresponds to a single order parameter dominating, making it essential to have an accurate free energy representation with one dominant order parameter. Additionally, for LCO, as the system evolves from disordered to ordered we expect for ordered domain of the material to only have one non-zero order parameter, making it important to understand the transition between ordered and disordered states along the path between $\eta_i=0$ to $\eta_i=0.5$, as shown by the black dotted line in Figure \ref{fig:wellsvertices2d}, where all other order parameters are close to 0. The procedure is shown in more detail in Algorithm \ref{alg: Sampling Vertices}.

\begin{algorithm}
\caption{Sampling Vertices Procedure}
\label{alg: Sampling Vertices}
\begin{algorithmic}[1]
\Procedure{SampleVertices}{rnd}
    \State Inputs: Perturbation magnitude($P$), Number of times to sample along each vertex ($N_v$), and the dimensions(dim). For LCO there are 7 dimensions, one corresponding to the composition of Lithium, and the other six correspond to order parameters. 
    \State Initialize $\eta_v$ as a zero matrix of size $(N_v \cdot \text{dim}, \text{dim})$
    \State Set $\eta_v[:, 0] \gets 0.5$ - This dimension corresponds to the composition of Lithium
    \For{$i = 1$ to dim$-1$}
        \State Find set of points corresponding to the $ith$ order parameter $I_i$ ie $I_i = [(i-1)*N_v:i*N_v]$
        \State Set $\eta_v[I_i,i]$ to something random between $\pm 0.5$
        \State set $\eta_v[I_i,j\neq i]$ = $P$ $\cdot$ rnd $= \epsilon$ 
        \State These dimensions correspond to order parameters, with $n_{w}[:,i] = \pm 0.5, i =1,..,$dim$-1$  being fully ordered
        \State An example row will look like $[0.5, \{-0.5,0.5\}, \epsilon_1,\epsilon_2,\epsilon_3,\epsilon_4,\epsilon_5 ]$. This mostly samples along the 1D vertex in $\eta_i$ where $\eta_0=0.5, \eta_j=0 (j \neq i)$. However, as the number of rounds increase, it allows for larger values for $\eta_j$. This ensures that some points are sampled along the 1D vertex itself, but allows for more variety in later rounds. 
    \EndFor

\EndProcedure
\end{algorithmic}
\end{algorithm}

We also sample at $\eta_0=0$ and $\eta_0=1$ because global sampling methods often struggle to adequately capture the behavior of the system near boundaries. These regions are typically underrepresented in the dataset, leading to sparse or noisy information for the machine learning (ML) models. Without explicitly including data from these endpoints, the ML models can become unstable or divergent near these boundary conditions, as they lack sufficient training data to accurately predict the system’s behavior. This can be seen by looking at the Billiard walk model in Figure \ref{fig:eta0eta1full}. Including these boundary points ensures that the model remains well-constrained across the entire range of $\eta_0$, improving its robustness and accuracy. Further details of this method are shown in Algorithm \ref{alg:endpoints}.

\begin{algorithm}
\caption{Sampling Endpoints Procedure}
\label{alg:endpoints}
\begin{algorithmic}[1]
\Procedure{SampleVertices}{rnd}
    \State Inputs: Perturbation magnitude($P$), Number of times to repeat the points at each endpoint ($N_e$), and the dimensions(dim)
    \State Initialize $\eta_e$ as a zero matrix of size $(2, \text{dim})$
    \State Set $\eta_e[1, 0] \gets 1$ - This row corresponds to a Li composition of $1$, and the first row corresponds to Li composition of $0$
    \State Repeat $\eta_e$ $N_e$ times along the axis
    \State Perturb $\eta_e$ by $P$ in all dimensions

\EndProcedure
\end{algorithmic}
\end{algorithm}

\subsubsection{High Error} \label{High Error theory}

High Error sampling is a common approach in many Active Learning and adaptive sampling methodologies \cite{tharwat2023survey,lookman2019active,eason2014adaptive,wu2023comprehensive,metta2021novel,benkert2023gaussian}. The rationale behind this method is grounded in several key assumptions. First, the primary goal is to develop a model that can predict points accurately across the input space. Points that exhibit high prediction errors are often located in sparsely populated or inherently difficult-to-model regions. These are precisely the regions where additional sampling is most beneficial, as it helps to refine the model’s accuracy in those areas. By focusing on points with High Error, the model can iteratively improve its predictions and reduce overall error. A simple example of High Error sampling is shown in Figure \ref{fig:error sensitivity}.

The implementation of High Error sampling is straightforward. For $j\in \{1,\dots, N\}$ data points, and dim is the dimension of each data point, the error is calculated using the formula 
\begin{equation}
    \text{MSE}_j = \sum_{i=1}^{\text{dim}} (\mu_{i_j}^{\text{MC}} -\mu_{i_j}^{\text{IDNN}})^2 
\end{equation}
 where $\mu_{i_j}^{\text{MC}}$ represents the true value or a more accurate estimate for dimension $i$ and data point $j$ and $\mu_{i_j}^{\text{IDNN}}$ represents the model's current prediction. Points are then sorted based on their calculated errors, and the algorithm perturbs those with the highest errors first. This targeted approach ensures that the High Error points are sampled early, leading to a more robust and accurate model over successive rounds.

\subsubsection{Highly Sensitive Regions} \label{senstivity theory}

Identifying regions where small changes in input parameters lead to significant variations in outputs indicates these regions are either not well modeled, or, if well-resolved, are have high sensitivity: They feature rapid changes in response where small changes in input \(\vec{\eta}\) result in large changes in output \(\vec{\mu}\). An example is shown by the area circled in Figure \ref{fig:error sensitivity}. Substantial errors result if such regions are not accurately captured. For phase field dynamics, inaccuracies in these regions can propagate, causing significant deviations in the simulated behavior of materials. These points can be identified by examining the eigenvalues of the Hessian matrix \(H\), where large eigenvalues indicate high sensitivity.

To systematically identify and address these regions, the IDNN is used to compute the Hessian matrix \(H\) of the free energy density function (equivalently, the gradient $\nabla_{\vec{\eta}} \vec{\mu}$) for each data point, which can be derived from both previously sampled data and predicted data points. This approach allows for a comprehensive search across a larger input space than  direct sampling would permit. By calculating the eigenvalues \(\vec{\lambda}\) of \(H\), we can determine the largest and smallest eigenvalues \(\lambda_{\text{max}}\) and $\lambda_\text{min}$ for each data point. Sorting these points by \(\lambda_{\text{max}}\) from highest to lowest prioritizes the most sensitive regions for additional sampling. Replicating these points on the basis of symmetry and perturbing them helps to refine the model further, ensuring that areas with high sensitivity are accurately resolved and reducing the likelihood of significant errors in the phase field dynamics simulations.

\subsubsection{Query by Committee } \label{QBC theory}

Query by Committee  (QBC) is a technique to identify regions or points within the input space where different models, trained on the same dataset but with different hyperparameters, exhibit significant disagreements. An example of a QBC problem is shown in Figure \ref{fig:qbc example}. This figure shows that with the same set of points, in some regions the fit can differ widely, and this is often due to complex regions or lack of points within that region, both of which would require additional sampling. The core idea behind QBC is to leverage the variability among these models to identify areas that are poorly represented or challenging to model, suggesting that additional sampling in these regions would be beneficial \cite{liu2018survey,fuhg2021state}.

As discussed in Section \ref{sec: hyperparameter}, a hyperparameter search is conducted in several rounds of the Active Learning workflow, resulting in several models that generate predictions for a given set of input points. For each of these models, we obtain a set of predictions, which allows us to calculate the average prediction across all models. By evaluating the fluctuation of predictions from this average, we can quantify the variability across different models. Specifically, we compute the fluctuation as the squared deviation from the average prediction, then sum these fluctuations across all dimensions. This aggregate measure of uncertainty highlights which points or regions have high variance among the models, indicating areas where the models disagree significantly.

Given  $N_\text{m}$ models, we first calculate the average for a prediction $y$.
\begin{equation}
    \vec{y}_{avg} = \frac{1}{h} \sum_{i=1}^{N_\text{m}} \vec{y}_i
\end{equation}
Then we find the fluctuation from the average
\begin{equation}
    \vec{y}_\text{fluc} = \frac{1}{h} \sum_{i=1}^{N_\text{m}} (\vec{y}_i -\vec{y}_{avg})^2
\end{equation}
 We then sum the fluctuation over all dimensions ($d$) to get the variance for each point ($\mathcal{V}$).
 
\begin{equation}
    \mathcal{V} = \sum_{j=1}^d(y_{fluc}^j)
\end{equation}

This fluctuation, across models and dimensions, is evaluated for each data point and is a measure of regions where model deviation is high. We then sort all variances from high to low and recommend the points with the largest variances for more sampling.

\subsubsection{Locating Free Energy Wells} \label{find wells theory}

Free energy density wells are of significant importance because they correspond to minimum energy crystal structures.  This can be seen in Figure \ref{fig:sample surface} at the regions circled in red. For instance in the case of LCO, the zigzag ordering shown in Figure \ref{fig:phasediagram} has a minimum where $\eta_0=0.5$ (half-lithiated) and one of the other order parameters is $\pm 0.5$ as can be seen in Figure \ref{fig:eta0eta1full}.

Furthermore, phase transitions from one crystal structure to another are governed by phase field dynamics. However, the  neighborhoods of energy wells in high-dimensional space often feature sharp gradients, making it challenging to model them accurately.

To identify and analyze these energy wells, we employ the Integrable Deep Neural Network (IDNN) model of the free energy density surface. The procedure begins with the computation of the Hessian matrix \( H \in \mathbb{R}^{d\times d}\) for each data point. This data can either come from previously sampled points or from predictions made by the IDNN. Although the IDNN allows for predictions over a larger input space than can be practically sampled, we utilize these predictions to enhance our exploration of free energy density landscapes $g(\vec{\eta})$.

Only those data points are retained where \( H \) is positive definite, which indicates local convexities characteristic of minima. This positivity is determined using the Cholesky decomposition   to \( \frac{1}{2}(H + H^T) \). Following this, the remaining data points are sorted based on the norm of their gradients, calculated as \( \sqrt{\sum_{i=0}^d \mu_i^2} \), with \( \mu_i \) representing the gradient components, recalling that $\mu_i = \partial g/\partial\eta_i$. Points exhibiting the smallest gradient norms are considered as corresponding to minima of the energy landscape.

By focusing on these criteria, we ensure a precise identification and analysis of energy wells, which is crucial for accurate modeling of phase transitions and material stability. This approach allows for a more thorough understanding of the energy landscape and enhances the effectiveness of phase field models.

\subsubsection{Regions of Instability} \label{regions of instability}

In LCO and other materials systems, the regions of interest include those with phase instability where the material is thermodynamically unstable such that it is likely to separate into two phases either through a nucleation and growth phenomenon or via spinodal decomposition \cite{CahnHilliard1958,Allen1979}. Proper representation of regions of phase instability in the free energy-order parameter space is therefore crucial for both  accurately simulating phase field dynamics as well as characterizing material properties. Additionally, since these regions are often relatively small in the high-dimensional space, they can prove difficult to resolve without targeted sampling. 

Phase instabilities are characterized by non-convexities of the free energy surface, so sampling near such regions is important for accurate model representation. An example of these non-convexities is the region circled in black in Figure \ref{fig:sample surface}. To achieve this, we use the IDNN to find the Hessian matrix \(H \) for each data point. These data points can be chosen from previously sampled data or from predicted data using the IDNN. The latter presents for exploration a larger input space than can be feasibly sampled. By finding the eigenvalues \(\lambda\) of \(H\), we can identify points where at least one eigenvalue, say \(\lambda_i\), is negative: (\(\lambda_i < 0\)). These points lie within regions of non-convexity. Once identified, these points can be replicated using symmetry with respect to the order parameters and perturbed to further refine the model's accuracy in these critical regions.

It is important to note that eigenvalues do not inherently correspond to specific order parameters. However, in the case of LCO, we observe that \(\lambda_0\) roughly aligns with the composition \(\eta_0\), while \(\lambda_1\) through \(\lambda_d\) approximately correspond to the order parameters \(\eta_1\) to \(\eta_d\). As such, we expect the normalized eigenvector \(\vec{\nu_0}\) to resemble \([1 \pm \epsilon, \epsilon, \epsilon, \epsilon, \epsilon, \epsilon, \epsilon]^\text{T}\), where \(\epsilon\) is a small value. Similarly, \(\vec{\nu_1}\) might look like \([\epsilon, 1 \pm \epsilon, \epsilon, \epsilon, \epsilon, \epsilon, \epsilon]^\text{T}\). It is also worth mentioning that the order of eigenvectors can vary depending on the mathematical toolkit used. In our results, the eigenvectors are returned in descending order of their corresponding eigenvalues (\(\lambda\)), from highest to lowest. If the order paremters \(\eta_1\) and \(\eta_2\) are swapped, the values of \(\vec{\nu}\) and \(\lambda\) remain unchanged. However, the correspondence between each eigenvector \(\vec{\nu}\) and its associated order parameter \(\eta\) will change, as each eigenvector will represent a different mode than it did before.

\subsubsection{Path of Lowest Free Energy} \label{LFE theory}

The path of lowest free energy is crucial for accurately simulating phase field dynamics in material systems. In phase field modeling, the evolution of microstructures is driven by minimizing the free energy of the system. This pathway determines how phases nucleate, grow, and transform over time, influencing the materials properties, while the dynamics determine its phase stability. By focusing on the lowest free energy path, simulations can access the most probable microstructural configurations and transitions between them, ensuring that the dynamic behavior of the material is realistically represented. Understanding and identifying these energy pathways is essential for predicting the material's response to external conditions, such as temperature, mechanical loads, composition, fluxes, currents and voltages, as well as their changes. Accurately modeling the free energy landscape, including identifying minima and transition states, allows for precise control over phase transformations and the design of materials with desired properties. Therefore, the path of lowest free energy serves as a fundamental guide for developing predictive and reliable phase field models in materials science. 

To determine the lowest free energy path we use the trained model to predict the free energy at various compositions of Li with different combinations of order parameters. For LCO we looked at compositions of Li between 0.45 and 0.55 at intervals of 0.002. At each composition of Li, $\eta_1, \eta_2$ are varied between 0 and 0.5 at intervals of 0.01 with the constraint that $\eta_1\geq \eta_2$ and the constraints from Equation \ref{eq: constraints on boundary}. We only considered two order parameters because from past works we have found that the lowest free energy neighborhoods occurs in regions where only two order parameters are nonzero at temperatures above 260 K. However, for different works this could be extended to be a search over all order parameters around the region of ordering.

\subsection{Hyperparameter Search} \label{sec: hyperparameter}
As the number of data points increases, the neural network architecture must be refactored to ensure sufficient representability. If the mean squared error (MSE) increases for two consecutive rounds this indicates the current architecture’s inability to capture the data’s complexity. In such cases, a random search over multiple sets of hyperparameters is conducted. These sets are used to train new models, and the one yielding the best performance is selected for future rounds. This approach dynamically adjusts the model’s complexity to scale with the data, maintaining accuracy and preventing performance degradation as new information is added.

The hyperparameters that we can search over include number of layers, neurons, activation function, dropout,  loss function, optimizer, learning rate, batch size, and learning rate reduction including patience (how long to wait before reducing the learning rate) and factor (how much should the learning rate be reduced by)

\subsection{Novelty} \label{sec: novelty}

An important aspect of determining optimal data points is ensuring that the new data points are novel, meaning they explore regions of the parameter space that have not been extensively sampled before. The novelty measure is designed to discourage exploration of already sampled areas. This approach helps to improve the efficiency of Active Learning by selecting data points that provide the most useful information for model refinement, avoiding redundant sampling, and focusing on areas that are underrepresented. For instance, after many rounds of the Active Learning workflow, the endpoints regions might be saturated and any additional information would be close to existing data points. 

The novelty process begins by calculating the distances between new candidate data points and the existing dataset. Using euclidean norms,  the average distance $\langle x \rangle$ to the k-nearest neighbors is computed for each new point. This provides a measure of how different or novel a point is relative to the existing data. Points that are too close to previously sampled points, as determined by the shortest distance metric, are assigned a novelty score of zero to avoid redundant data collection.  Then all points with a novelty score of zero or having an otherwise low score are not sampled. This ensures that the algorithm favors new regions of the parameter space over regions that have already been explored.

Additionally, the model predicts key outputs, such as free energy, chemical potential ($\mu$), and the Hessian matrix for each new point. The Hessian matrix, representing the curvature of the model’s predictions, is used to calculate the eigenvalues and eigenvectors. The largest absolute eigenvalue,  an indicator of the local sensitivity (Section \ref{senstivity theory}), is also a driver of uncertainty in the model’s predictions, if not accurately evaluated. This maximum eigenvalue is combined with the distance-based novelty score, providing a final score that reflects both the spatial novelty of the point and its importance based on the model’s prediction uncertainty.

By combining these two factors—distance and uncertainty—the novelty process ensures that selected data points are not only new but are also likely to contribute significant information for improving the model. The novelty score is calculated for each point:
\begin{equation}
 s_i = \lambda_{i} *\langle x_i \rangle
\end{equation}
Then the points are shuffled. The algorithm iterates over the set of normalized scores \( \{s_i\} \) and selects points based on a comparison criterion. A point \( p_i \) is retained if:  

\[
s_i > s_{i-1} \quad \text{or} \quad s_i > r \quad \text{and} \quad s_i \neq 0
\]

\noindent where \( s_{i-1} \) is the score of the previously selected point, and $r$ is a randomly sampled threshold. Depending on input it can either be $r_1$ or $r_2$ where $r_2$ is more aggressive and removes more points. 

\[
r_1 = \xi \cdot \frac{1}{n} \sum_{i=1}^{n} s_i, \quad \xi \sim U(0,1)
\]
\[
r_2 = \xi \cdot \max( s_i), \quad \xi \sim U(0,1)
\]

Here, \( \xi \) is a uniform random variable, ensuring diversity in selected points. $n$ is the number of points recommended within a round of the Active Learning workflow. The max is the maximum novelty that has been recommended in that particular round. This method helps to maximize the effectiveness of data collection in iterative learning processes, balancing exploration of new regions with areas where the model may require more refinement.

\subsection{Dynamic Sampling}

In Active Learning, it is important to adapt the data sampling strategy based on the evolving performance of the model. This ensures that the most informative data points are selected for each round, improving the overall efficiency of the learning process. A key aspect of this adaptation is dynamically reweighting different criteria according to how much the model improves in specific areas after incorporating new data. By doing this, the Active Learning process can allocate more importance to areas where improvements are slower, ensuring that diverse regions of the parameter space are explored. For instance, if we are sampling within the wells using the Sampling Wells criterion, but the model is already highly accurate in the wells, it is of limited information gain to continue sampling there, in favor of other regions.

The dynamic reweighting process works by first evaluating the performance of previously trained models and the current one. For each criterion, such as High Error (Section \ref{High Error theory}), sensitivity, or the finding wells, the model calculates the mean squared error (MSE) between predicted and actual values of key quantities, like the chemical potential ($\mu$). This is done for both the previous model rounds and the current one, allowing for a comparison of their respective performances across  data points sampled from different regions.

The improvement for each criterion is determined by the difference between the MSE of previous model rounds and the current one. Positive improvements indicate that the model has better learned to predict the region in question, while no improvement or negative values signal areas where the new data points did not provide useful information for training the model. The weights for each criterion are adjusted based on the magnitude of improvement. Criteria that show greater improvement (ie the largest increase in MSE in those regions) are rewarded with higher weights, encouraging the selection of data points from those criteria in subsequent rounds.

Additionally, checks are implemented to prevent the weight for any category from falling below zero, ensuring that every criterion maintains some influence in the Active Learning process. Once the weights are updated, they are normalized so the average value of the weights is 1, maintaining a roughly equal number of points sampled in each Active Learning round. 

This reweighting approach is useful because it allows the Active Learning algorithm to focus on criteria under which the model improves by using them  to quantitatively drive the sampling, providing an automated mechanism to balance exploration and exploitation. It promotes a more targeted data sampling strategy, which ultimately accelerates convergence by focusing the learning on criteria, and through them, on areas that need improvement, while still considering other important factors. This adaptive strategy allows the model to become more efficient over time, resulting in a more accurate and comprehensive representation of the data.

\section{Results} \label{sec:results}
\begin{figure}[h!]
    \centering
    \begin{subfigure}{0.48\linewidth}
        \centering
        \includegraphics[width=\linewidth]{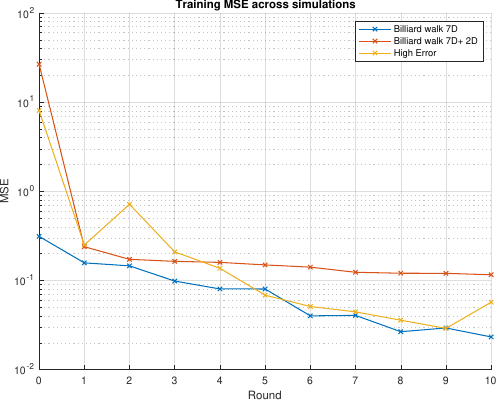}
        \caption{}
        \label{fig:training mse1}
    \end{subfigure}
     \hfill
    \begin{subfigure}{0.48\linewidth}
        \centering
        \includegraphics[width=\linewidth]{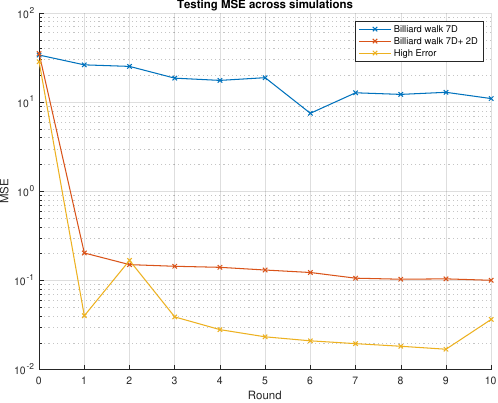}
        \caption{}
        \label{fig:testing mse1}
    \end{subfigure}
    \caption{The results of the Active Learning workflow for Billiard walk 7D sampling only (Billiard walk 7D), Billiard walk sampling 7D and 2D (Billiard walk 7D + 2D) and Billiard walk sampling 7D and 2D with High Error points (High Error). Mean Squared Error (MSE) vs. round for (a) the final training set (from round 10) and (b) the testing set. High Error has a lower MSE for testing compared to training due to the choice for data in the testing set as discussed in the text.}
    \label{fig:mse_comparison1}
\end{figure}

\begin{figure}[h!]
    \centering
    \begin{subfigure}{0.48\linewidth}
        \centering
        \includegraphics[width=\linewidth]{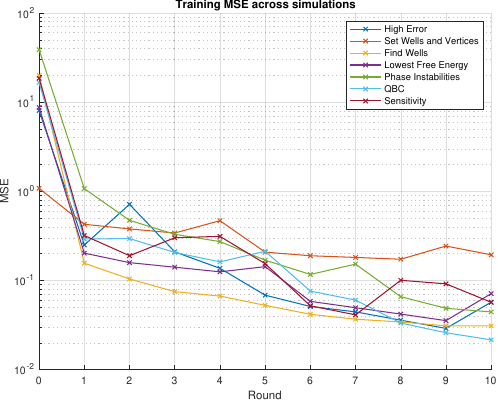}
        \caption{}
        \label{fig:training mse2}
    \end{subfigure}
    \hfill
    \begin{subfigure}{0.48\linewidth}
        \centering
        \includegraphics[width=\linewidth]{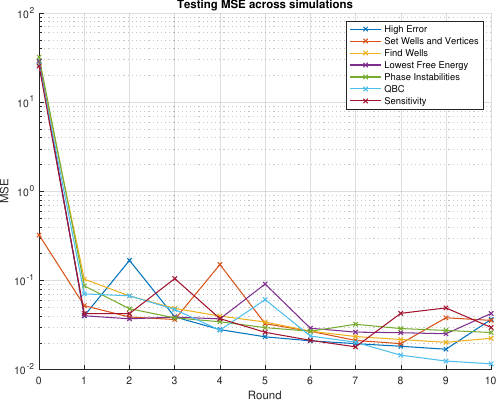}
        \caption{}
        \label{fig:testing mse2}
    \end{subfigure}
    \caption{The results of the Active Learning workflow based on different criteria are shown. All results use Billiard walk sampling in 7D and 2D with high-error points. The High Error line is the same as in Figure \ref{fig:mse_comparison1}, while the other lines incorporate the additional criteria specified by their names. Mean Squared Error (MSE) vs. round for (a) the final training set (from round 10) and (b) the testing set. QBC has a lower MSE for testing compared to training due to the choice for data in the testing set as discussed in the text.}
    \label{fig:mse_comparison2}
\end{figure}

\begin{table}[ht]
    \centering
    \tiny
    \renewcommand{\arraystretch}{1.2} 
    \begin{tabular}{|l|c|c|c|c|c|c|c|c|}
    \hline
    \textbf{Method} & \textbf{Training MSE} & \textbf{Testing MSE} & \textbf{BW\_2D} & \textbf{BW} & \textbf{LFE} & \textbf{Non-Convexities} & \textbf{Vertices} & \textbf{Wells} \\
    \hline
Billiardwalk        & 0.002329 & 1.096225 & 4.010211 & 4.834101 & 0.001974 & 0.001675 & 2.731979 & 0.145381 \\
Billiardwalk2D      & 0.011597 & 0.010057 & 0.016334 & 0.013783 & 0.009695 & 0.007169 & 0.006749 & 0.014415 \\
Find Wells          & 0.003101 & 0.002252 & 0.003503 & 0.001553 & 0.002521 & 0.002004 & 0.001142 & 0.000951 \\
Set Wells/Vertices  & 0.017319 & 0.001960 & 0.002145 & 0.002839 & 0.002027 & 0.001022 & 0.000996 & 0.000192 \\
Lowest Free Energy  & 0.003550 & 0.002529 & 0.002939 & 0.004474 & 0.002269 & 0.002201 & 0.001994 & 0.000603 \\
Phase Instabilities & 0.004443 & 0.002612 & 0.004803 & 0.005583 & 0.002235 & 0.001044 & 0.002345 & 0.000255 \\
QBC                 & 0.002160 & 0.001164 & 0.000760 & 0.001551 & 0.001197 & 0.000554 & 0.000527 & 0.000476 \\
High Error          & 0.002905 & 0.001696 & 0.001217 & 0.002252 & 0.001678 & 0.002523 & 0.001187 & 0.002415 \\
Sensitivity         & 0.004122 & 0.001804 & 0.001662 & 0.002342 & 0.001853 & 0.000973 & 0.000991 & 0.000795 \\
allCriteria1        & 0.003185 & 0.003096 & 0.005665 & 0.006713 & 0.002541 & 0.001263 & 0.004186 & 0.001198 \\
allCriteria2        & 0.003252 & 0.002662 & 0.004325 & 0.003999 & 0.002560 & 0.001799 & 0.002177 & 0.003484 \\
   \hline
    \end{tabular}
    \caption{Performance metrics for various methods across different simulation tasks.}
    \label{tab:performance}
\end{table}

\begin{figure}
    \centering
    \includegraphics[width=\linewidth]{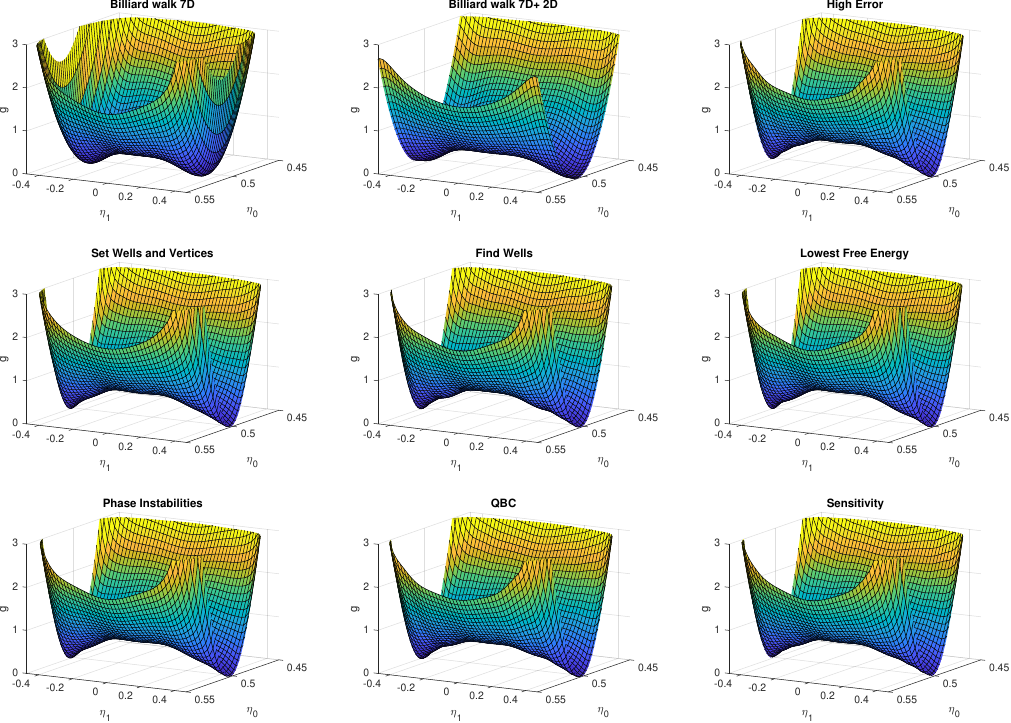}
    \caption{The free energy density function as a function of composition($\eta_0$) and one order parameter($\eta_1$) where the other order parameters: $\eta_j=0$ for $j=2,...,6$. The composition is constrained by $0.45 \leq \eta_0 \leq 0.55$, such that it is focused near orderings.}
    \label{fig:eta0eta1closeup}
\end{figure}

\begin{figure}
    \centering
    \includegraphics[width=\linewidth]{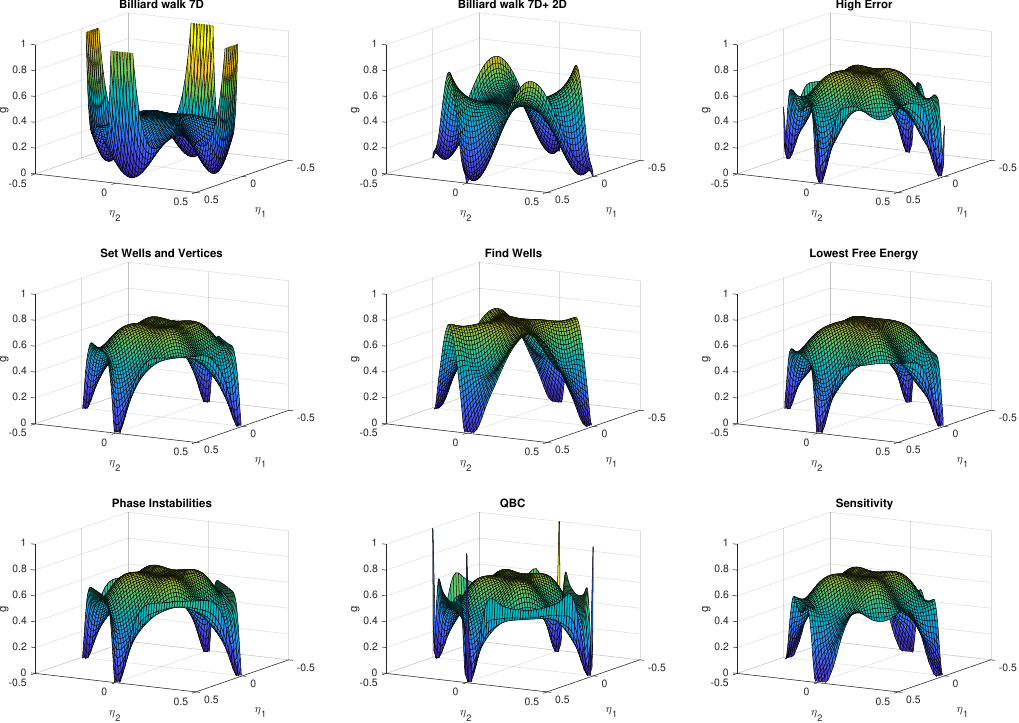}
    \caption{The free energy density function as a function of the order parameters $\eta_1,\eta_2$ where the composition $\eta_0=0.5$ and the  order parameters $\eta_j=0$ for $j=3,...,6$}
    \label{fig:eta1eta2}
\end{figure}

\begin{figure}
    \centering
    \includegraphics[width=\linewidth]{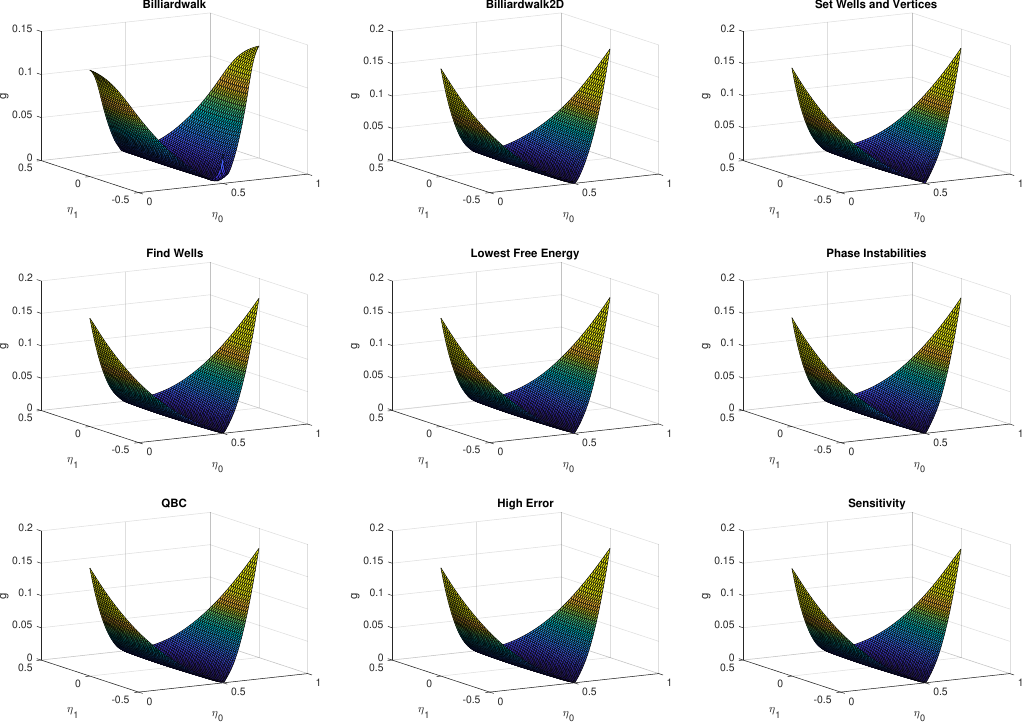}
    \caption{The free energy density function as a function of composition($\eta_0$) and one order parameter($\eta_1$) where the other order parameters: $\eta_j=0$ for $j=2,...,6$. The composition is unconstrained. }
    \label{fig:eta0eta1full}
\end{figure}

\begin{figure}[h!]
    \centering
    \begin{subfigure}{0.48\linewidth}
        \centering
        \includegraphics[width=\linewidth]{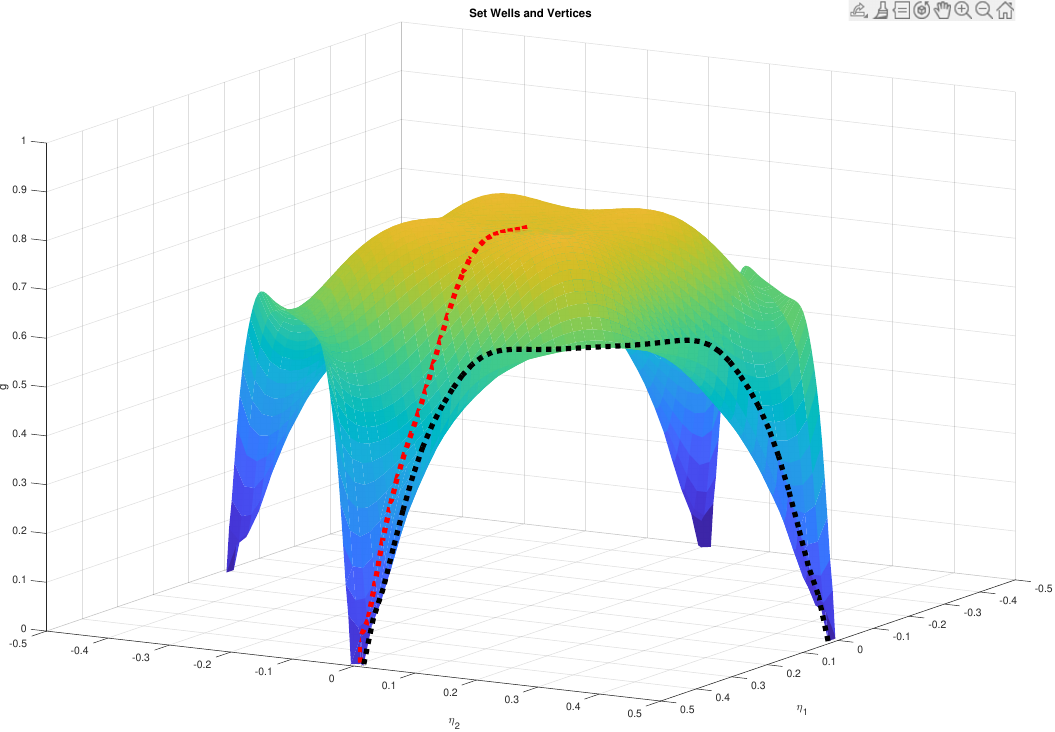}
        \caption{}
        \label{fig:wellsvertices2d surf}
    \end{subfigure}
    \hfill
    \begin{subfigure}{0.48\linewidth}
        \centering
        \includegraphics[width=\linewidth]{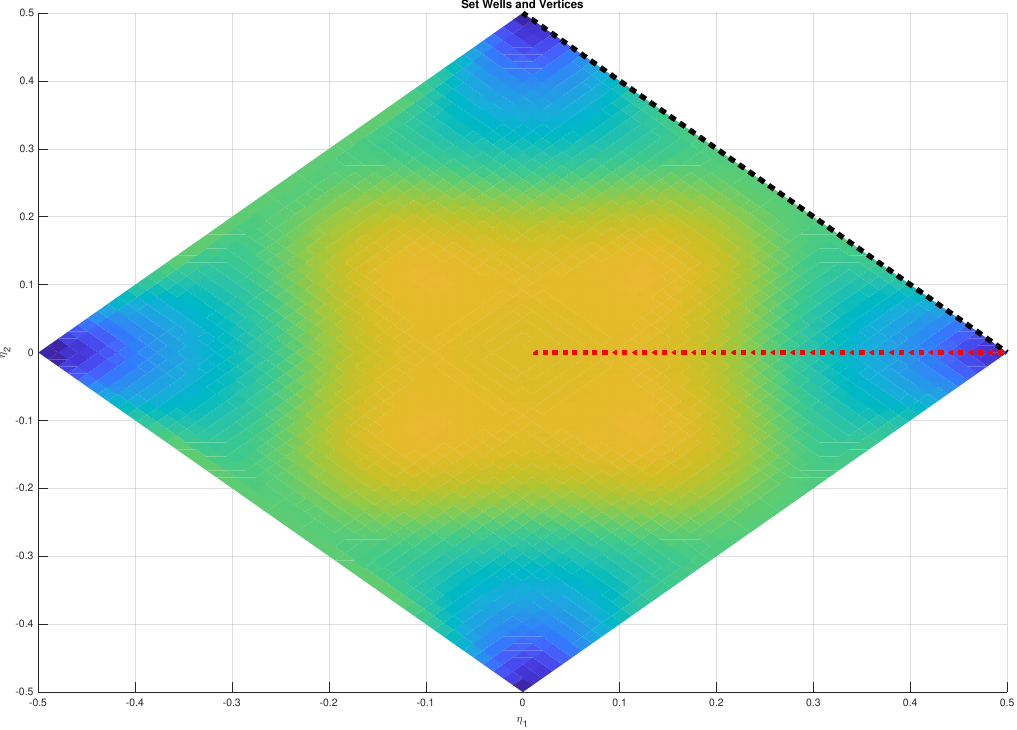}
        \caption{}
        \label{fig:wellsvertices2d plane}
    \end{subfigure}
    \caption{The free energy density function for the Set Wells and Vertices Method as a function of the order parameters $\eta_1,\eta_2$ where the composition $\eta_0=0.5$ and the order parameters $\eta_j=0$ for $j=3,...,6$. The red dotted line shows the order to disordered free energy path and the black dotted line shows an ordered to ordered energy transition. Figures (a) and (b) show the same data but (b) shows the $\eta_1-\eta_2$ plane.}
    \label{fig:wellsvertices2d}
\end{figure}

\subsection{Individual Criteria}

We performed an Active Learning workflow for each criterion with 10 rounds. Dynamic sampling and novelty sampling are not applicable to testing the criteria individually. For each criterion, we first ran the Active Learning workflow several times without a hyperparameter search, varying the number of points to sample with that criterion and varying how many time to replicate each point for perturbation of values.  To avoid a high computation cost we trained a NN on Monte Carlo data, and we used this surrogate Monte Carlo to give us the $(\vec{\eta},\vec{\mu})$ pairs.  This method provides insight to the optimal number of points, and all results discussed below were obtained with this method. We then run the Active Learning workflow with the real MCMC calculations and hyperparameter search.

For each simulation, points were first obtained by the Billiard walk. We did one sampling called Billiard walk 7D  with 1000 points $\vec{\eta}\in \mathbb{R}^7$.  For every other simulation our Billiard walk sampling had 950 points $\vec{\eta}\in \mathbb{R}^7$ and 50 points $\langle \eta_0,\eta_1,0,0,0,0,0\rangle^\text{T}$, which we refer to as Billiard walk-2D. We tried a few ratios of the number of 7D points to 2D points ($0\%,5\%,20\%$) and found that the workflow with $5\%$ 2D points had the lowest final testing MSE. Therefore, we used that ratio in our other simulations. Billiard walk-2D just used the 950 7D points and 50 2D points. Then the High Error criterion method used Billiard walk sampling and 1200 High Error points (the top 400 highest error points, triplicated and perturbed in each $\eta_i, \; i = 0,\dots 6$ by $\pm 0.075$.  

The rest of the criteria used the same setup of High Error sampling supplementing Billiard walk (950 points in 7D and 50 in 2D) sampling as discussed above in addition to the sampling of each criterion. Each criterion was run with the surrogate MC with the number of points for that criterion varied to determine the value that yields the lowest loss. Then each was run with the actual Monte Carlo sampling.  The global Billiard walk sampling points were the same for each Active Learning Run we did. This was because global sampling is not dependent on any training results, and minimzied the number of calculations we need to make while also providing a degree of standardization.

We determine the MSE for the model from each round with respect to the training data for all 10 rounds for that particular criterion. Additionally we determine the MSE with respect to a testing data. The testing data was generated using Billiard walk sampling (7D and 2D), the  regions of interests (wells, vertices, and end points), phase instabilities, and the lowest free energy. This achieves sufficient coverage of the space in $\mathbb{R}^7$. 

We have plotted the losses per round in Figures \ref{fig:training mse1} and \ref{fig:testing mse1} for Billiard walk 7D, Billiard walk 7D+2D, and Billiard walk 7D+2D + High Error. The Billiard walk-only sampling had the highest testing MSE, which is expected. The Billiard walk-2D has significant improvement--almost 2 orders of magnitude--over Billiard walk only with 7D as seen by looking in Fig \ref{fig:testing mse1}. However, both Billiard walk versions  do not show significant improvement after round 1. Then,  adding the High Error points improves the testing MSE by another order of magnitude.

The training MSE shown in Figure \ref{fig:training mse1} show that the MSE decreases over time suggesting that the our neural networks are fitting well to the data points they are trained on. Billiard walk has the lowest MSE because it does not have many points in difficult to capture regions like orderings, endpoints or wells. Therefore the IDNN has an easier time training to those data points. 

For the other criterion, which all include Billiard walk 7D+2D + High Error and the additional criteria, those results are shown in Figures \ref{fig:training mse2} and \ref{fig:testing mse2}.Comparison across the remaining criteria which were used in addition to the Billiard walk plus High Error, gets more complicated especially since many of the methods  target specific regions of interest, and are not designed to minimize the MSE across all the sampling points. They all show the same general trends and the testing MSE are mostly of a similar magnitude to the Billiard walk 7D+2D + High Error results. Overall, the best results for the testing data MSE was found using Query by Committee . The training MSE in Figure \ref{fig:training mse2} show that the model improves with each round and more data, but the training MSE was starting to flatten, so the additional points were not adding much more value. Several methods including High Error and Query by Committee have a lower MSE for testing data versus training data. This is due to the specific data set used for testing which emphasizes the physically meaningful regions such as wells and phase instabilities, and does not include some of the harder to capture points from criteria including High Error and QBC.

For each criterion, the loss was further broken down for the round that gives the lowest error for the training data across all 10 rounds. It shows how well each criterion performs over the entire space, and in the different regions of interest. The results for this are shown in Table \ref{tab:performance}. Here the columns aside from Training MSE and Testing MSE represent a set of points corresponding to different methods. Each row shows the results from different Active Learning runs using those criterian. The allCriteria1 and allCriteria2 use all methods with allCriteria2 additional using Dynamic Sampling to reweight the relative sampling of different methods.

Overall, the best results for the testing data MSE was found using Query by Committee . It had consistently low losses throughout  the rounds in Figure \ref{fig:training mse2}, and had the lowest or among the lowest training MSE across all methods in the table \ref{tab:performance}. Additionally, the sensitivity method performs fairly well.

Setting wells/vertices also yields low MSE's overall and yielded the lowest testing MSE in the wells regions. However, it is more volatile than Query by Committee, with the training MSE changing significantly between rounds. Setting wells/vertices works better than finding wells for lowering the MSE with regard to non convexities, Sampling Vertices and Sampling Wells. However, finding wells works better otherwise. This is likely because other apparent ``wells'' are identified in the free energy space, that satisfy the Find Wells criterion, causing  points to be added there. However, rather than real wells, these are regions of  non-smoothness due to a poor neural network fit.
 
The lowest free energy method had the lowest testing MSE along the lowest free energy path, but overall had a relatively high MSE, since it is targeting a highly specific area. 

For the general error reduction techniques: sensitivity and Query by Committee  both sample predominantly around the regions of ordering and phase instability $0.4 \le \eta_0 \le 0.6$ with two nonzero order parameters $\eta_i,\eta_j$, $i,j = 1,\dots 6$, $i \neq j$. The High Error method also predominantly samples within that region - but samples additionally in $\eta_0 > 0.7$ and $\eta_0 < 0.3$. 

The non-convexities method samples predominantly in a few regions, outside of the regions of phase instability that we might expect around the ordering. For instance, a lot of points sampled had a composition of Li less than $0.4$  In this work, the final free energy representation does not have these (or any) regions as non-convex (ie looking like \ref{fig:sample surface}) suggesting this method worked primarily to identify mistaken phase instability predictions by the neural network, to sample additionally in those regions, and then to correct the network.   

The lowest free energy criterion is only set to sample in $0.45\le \eta_0 \le 0.55$. For the first few rounds, the largest order parameter, $\eta_i$, $i \in \{1,\dots,6\}$ can be 0 (suggesting completely disordered) or close to 0.5 (suggesting completely ordered). Eventually as the neural network converges, it is always ordered around 0.5, $\eta_i = \pm 0.5$ as is dictated by the physics.

The Find Wells method works as expected, with most points having composition $\eta_0 \approx 0.5$ and around $\eta_1 \approx 0$ or $\eta_1>0.2$.

We analyzed the free energy landscapes generated by various Active Learning and sampling strategies, comparing their key features and identifying unique behaviors. Below, we summarize the results for each method, focusing on the space around ordering. Using symmetry with respect to $\eta_i, i \in \{1,\dots, 6\}$, we focus on $\eta_0,\eta_1$ varying while $\eta_2,...,\eta_6=0$  and on at $\eta_1,\eta_2$ varying while $\eta_0=0.5$ and $\eta_j = 0, j \in \{3,\dots 6\}$.

The free energy surfaces look fairly similar in the slice of space over which $\eta_0,\eta_1$ vary while $\eta_2,...,\eta_6=0$ as can be seen in Figure \ref{fig:eta0eta1closeup} and \ref{fig:eta0eta1full}. Aside from the Billiard walk, all the criteria show the same general form, which is fairly smooth, except that many diverge in the neighborhood of $\eta_1=0.5$. They can diverge either to $+\infty$ (e.g., in the case of “Find Wells”) or to $0\infty$  (e.g., in “Set Wells and Vertices”). These divergences are significant. If the divergence is toward $+\infty$ , the global minimum—representing the energy wells—will occur near $\eta_1=\pm 0.5$, with $|\eta_1| <0.5$. However, if the divergence is toward $-\infty$ , this poses a challenge for phase-field calculations. In such cases, the phase field evolves toward $\eta_1=0.5$, but it becomes trapped in the well because the energy is spuriously lowered, misrepresenting the true surface.

The learnt representations get more complicated if $\eta_1,\eta_2$ are varies while $\eta_0=0.5$, that is near the ordered region on the $\eta_0$ axis. The results are shown in Figure \ref{fig:eta1eta2}. All versions, expect for Billiard walk sampling alone, show the minimum energy (ignoring divergence) occurs when $\eta_0=0.5$ and close to $\eta_1$ or $\eta_2=\pm 0.5$ as expected. This is because the zigzag ordering has the lowest formation energy as shown from DFT results, and the crystal is ordered with the zigzag ordering at $\eta_0=0.5,\eta_i=\pm 0.5,i=1,..,6$. 

Most criteria additionally show a local minimum at $\eta_0=0.5, \eta_i=0 (i=1,..6)$, which suggests that the completely disordered state is energetically favored. These minima are most prominent for the criteria which have the lowest MSE for Sampling Wells, including Query by Committee, Set Wells/Vertices,Phase Instabilities, and Lowest Free Energy. It is notable, that around this local well, as can be seen in Figure \ref{fig:eta1eta2} most criteria show that there are local maxima where both $\eta_1$ and $\eta_2=0.1$ versus only $\eta_1=0.1$ and $\eta_2=0$. In contrast, for the lowest free energy criterion, the surface appears to be locally spherically symmetric around $\eta_1,\eta_2=0$. This is important for understanding how the phase field dynamics will evolve. Most criteria will show an evolution from disordered towards one order parameter, ie $\eta_1$ or $\eta_2$ as shown in the Figure. However, the lowest free energy version could evolve towards a combination of order parameters, which would show a very different phase surface. We also note in passing that this local minimum is observed only for the MC  data at $260$K. Atsampling temperatures $300-340$K  the ordered state is the lowest energy configuration.

Additionally, there is significant variation along the boundary between $\eta_1\sim 0.4, \eta_2=0$ through $\eta_1\sim  0.25, \eta_2\sim 0.25$ to $\eta_1=0,\eta_2\sim 0.4$. Some criteria show that along that path there is a local minimum at $\eta_1\sim  0.25, \eta_2\sim 0.25$, while others show that there is a maximum there. Most criteria show that the energy along this path is significantly higher than the ordered region, except for the Find Wells criterion which does not show a large increase in free energy along that path as compared to the fully ordered regime. This path is important in phase field dynamics, because if there are two ordered regions with different order parameters next to each other, they will want to merge to lower the interfacial energy. Understanding if this transformation from one order parameter to another is possible and how it occurs is vitally important, this is the block dotted line along the free energy surface shown in Figure \ref{fig:wellsvertices2d}.

\subsection{Additional Active Learning Methods}
\label{sec:addAL}

\begin{figure}[h!]
    \centering
    \begin{subfigure}{0.48\linewidth}
        \centering
        \includegraphics[width=\linewidth]{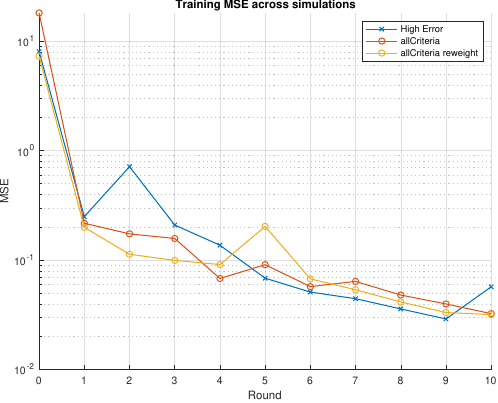}
        \caption{}
        \label{fig:training mse3}
    \end{subfigure}
    \hfill
    \begin{subfigure}{0.48\linewidth}
        \centering
        \includegraphics[width=\linewidth]{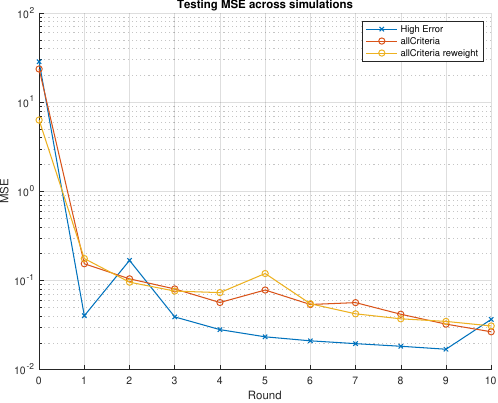}
        \caption{}
        \label{fig:testing mse3}
    \end{subfigure}
    \caption{This figure shows the Active Learning results for High Error as a comparison point, where High Error consists of Billiard walk 7D + 2D and High Error and is also shown in Figures \ref{fig:mse_comparison1},\ref{fig:mse_comparison2}. Then the results for allCriteria are shown with and without dynamic reweighting. Mean Squared Error (MSE) vs. round for (a) the final training set (from round 10) and (b) the testing set.}
    \label{fig:mse_comparison3}
\end{figure}

\begin{figure}[h!]
    \centering
    \begin{subfigure}{0.48\linewidth}
        \centering
        \includegraphics[width=\linewidth]{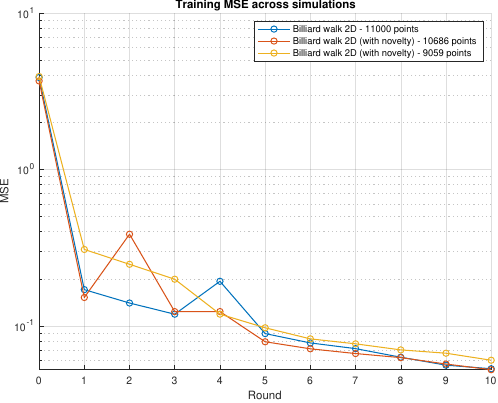}
        \caption{}
        \label{fig:training mse4}
    \end{subfigure}
    \hfill
    \begin{subfigure}{0.48\linewidth}
        \centering
        \includegraphics[width=\linewidth]{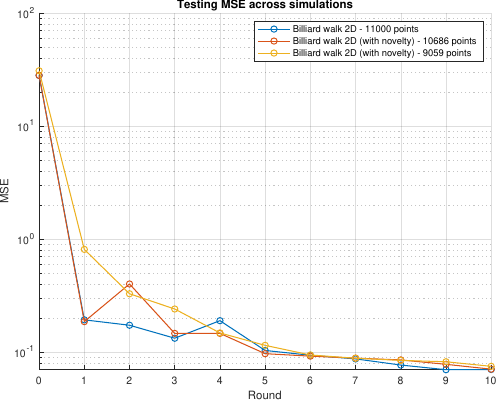}
        \caption{}
        \label{fig:testing mse4}
    \end{subfigure}
    \caption{This figure shows the Active Learning results for Billiardwalk 2D with and without novelty. The novelty with fewer points has a stricter requirement to keep points. Mean Squared Error (MSE) vs. round for (a) the final training set (from round 10) and (b) the testing set.}
    \label{fig:mse_comparison4}
\end{figure}

\begin{table}[ht]
    \centering
    \tiny
    \renewcommand{\arraystretch}{1.2} 
    \begin{tabular}{|l|c|c|c|c|c|c|c|c|}
    \hline
    \textbf{Method} (all Billiard walk 2D) & \textbf{Training MSE} & \textbf{Testing MSE} & \textbf{BW\_2D} & \textbf{BW} & \textbf{LFE} & \textbf{Non-Convexities} & \textbf{Vertices} & \textbf{Wells} \\
    \hline
Original- 11000 points      &                0.005329     &  0.007029  &        0.018108    &   0.019614     &        0.005087     &     0.005339    &      0.005489   &    0.008806 \\
With novelty $r_1$ - 10686 points    &    0.005255     &  0.007075   &       0.019083   &    0.019103       &      0.004985   &       0.005840     &     0.006352   &    0.008879 \\
With novelty $r_2$ - 9059 points      &   0.006040     &  0.007498     &     0.024098    &   0.018102   &          0.005605   &       0.005690       &   0.006142    &   0.007199 \\ 
   \hline
    \end{tabular}
    \caption{MSE for Novelty Sampling }
    \label{tab:performance novelty}
\end{table}

We additionally ran the Active Learning Workflow with all criteria. We began with 2000 global points (95$\%$ 7D and 5$\%$ 2D) and 20 points for each other criterian. This was chosen by running simulations with a surrogate model with various numbers of points, and this combination was found to yield the lowest MSE. Run with real CASM data, this performed comparable to some of the individual criterion methods but not as well as the best as can be seen in Table \ref{tab:performance}. We then ran this same method, but allowed for reweighting with and $\alpha=10000$. Again, this $\alpha$ was chosen because using the surrogate model it yielded the lowest MSE. The results for this are shown in  Table\ref{tab:performance}. This automatic reweighting caused the largest increase in points for Sampling Wells and High Error. This yielded a lower MSE for testing data in the wells or non-convex regions and yieded a comparable testing MSE overall as seen in Figure \ref{fig:testing mse3}. This shows this type of reweighting can provide valuable insight into difficult to capture regions, but can also lead to overfitting of these regions. 

To test our novelty method, we ran the Active Learning through 10 rounds, with 1000 global points  (95$\%$ 7D and 5$\%$ 2D)  for each round ( beginning at round 0). We also did not use a hyperparameter search, since we were just comparing billiardwalk with and without novelty constraints. Without any novelty constraints 11000 points were sampled. We then did novelty constraints with $r_1$ from Sec \ref{sec: novelty}, where 10686 points were sampled, and with $r_2$ where 9059 points were sampled. The results are shown in Figures \ref{fig:testing mse4}
 and \ref{fig:training mse4}. As can be seen, while the 2nd novelty sampling has a slightly higher testing and training MSE in the first few rounds, by the 10th round the MSEs were all comparable. Additionally, the MSE's in specific regions are also comparable for all three runs as seen in Table \ref{tab:performance novelty}, this is despite the $r_2$ Active Learning run having about $20\%$ fewer points.

\section{Discussion}

In the full seven-dimensional space, Billiard walk sampling struggles to adequately cover key regions of interest, particularly in areas that require attention to ordering. This limitation drove the addition of 2D Billiard walk  to improve sampling in these crucial regions. High Error sampling, while effective at refining the model, depends on previously sampled points, meaning it is constrained by the other sampling criteria in use.

Sampling within wells provides valuable insights to their shape; however, it does not reveal sufficient information about the relative depths of wells. During training, excessive sampling within the wells was shown to reduce the MSE between the predictions and labels for $\vec{\mu}$. However, this occasionally resulted in a well corresponding to a disordered crystal ($\eta_0 = 0.5, \eta_i = 0$) appearing as the global minimum, rather than the well at the ordered regions ($\eta_0 = 0.5, \eta_i = 0.5$), potentially resulting in a model that seems well-fitted but inaccurately represents material phenomena. To avoid this, sampling must include points along the vertices that define the free energy surface between wells, covering the entire region where $\eta_0 = 0.5$ rather than focusing solely on the wells themselves. We also reiterate that this minimum appears only at $260$K, and disappears between $300$K and $340$K.

Standard error-reduction techniques, such as High Error, sensitivity, and Query by Committee , demonstrate strong performance across various regions, maintaining consistency in their results. Conversely, physics-based sampling methods—like locating wells, identifying phase instabilities, and mapping free energy paths—perform well within the specific regions they target but may lack the versatility to generalize as effectively across the broader sampling space. Effectively, these criteria promote overfitting to certain geometric features of the free energy landscape.

By the 10th round, the training and testing curves have largely stabilized. Overall, the MSE curve for the training set appears much smoother than that of the testing set, as expected.

\section{Conclusion}

In this work, we have introduced an adaptable Active Learning workflow for training a free energy model using derivative data derived from DFT-informed Monte Carlo simulations. Central to this approach are the data sampling techniques employed to identify which Monte Carlo points to sample within the 7D order parameter space. These techniques include a combination of space-filling, error-driven, and physics-driven methods. While the space-filling and error-driven approaches enhance generalizability across the entire 7D space, the physics-driven methods provide valuable insights into geometrically significant regions, albeit at the expense of broader applicability.

We have also implemented strategies to optimize the overall workflow, including neural network hyperparameter searches, dynamic reweighting of sampling criteria, novelty enforcement, and stopping criteria. These enhancements are critical for balancing model performance and computational efficiency. For instance, hyperparameter searches ensure that the neural network architecture evolves appropriately as more data points become available, preventing issues of overfitting or underfitting. Dynamic reweighting can yield lower errors in difficult to capture regions such as wells or non-convexities, but can cause overfitting. 

The applicability and effectiveness of this workflow depend on several key factors, including the level of prior knowledge about the material system, the computational cost of the Monte Carlo simulations, and whether the objective of the free energy model prioritizes generalizability or high accuracy in physically significant regions. By tailoring the methods to these considerations, this workflow provides a robust framework for efficiently and effectively training free energy models.

\bibliographystyle{unsrt}
\bibliography{references}

\end{document}